\documentclass[11pt,a4paper]{article}
\pdfoutput=1
\usepackage{jheppub}
\usepackage[utf8]{inputenc}
\usepackage{dcolumn}
\usepackage{bm,amsfonts,amsthm,amsmath,amssymb}
\usepackage{graphicx}
\usepackage{subfigure}

\newcommand{\bi}{\begin{itemize}}
\newcommand{\ei}{\end{itemize}}

\newcommand{\be}{\begin{equation}}
\newcommand{\ee}{\end{equation}}
\newcommand{\nn}{\nonumber}

\newcommand{\Ncal}{{\cal N}}
\newcommand{\Ocal}{{\cal O}}
\newcommand{\Pcal}{{\cal P}}
\newcommand{\Klein}[2]{ \ensuremath{ \left( #1 , #2\right)_{{\rm KG}}}}
\newcommand{\comm}[2]{ \ensuremath{ \left[ #1 , #2\right]}}
\newcommand{\sech}{{\rm sech}}
\newcommand{\Real}{{\rm Re}}

\newcommand{\alphaN}{\ensuremath{\alpha_{{\rm N}}}}
\newcommand{\betaN}{\ensuremath{\beta_{{\rm N}}}}
\newcommand{\alphaL}{\ensuremath{\alpha^{{\rm L}}}}
\newcommand{\alphaLt}{\ensuremath{\tilde{\alpha}^{{\rm L}}}}
\newcommand{\betaL}{\ensuremath{\beta^{{\rm L}}}}
\newcommand{\alphaH}{\ensuremath{\alpha^{{\rm H}}}}
\newcommand{\betaH}{\ensuremath{\beta^{{\rm H}}}}
\newcommand{\alphaLN}{\ensuremath{\alpha^{{\rm L}}_{{\rm N}}}}
\newcommand{\betaLN}{\ensuremath{\beta^{{\rm L}}_{{\rm N}}}}
\newcommand{\alphaLNt}{\ensuremath{\tilde{\alpha}^{{\rm L}}_{{\rm N}}}}
\newcommand{\betaLNt}{\ensuremath{\tilde{\beta}^{{\rm L}}_{{\rm N}}}}
\newcommand{\alphaML}{\ensuremath{\alpha^{{\rm ML}}}}
\newcommand{\betaML}{\ensuremath{\beta^{{\rm ML}}}}
\newcommand{\alphaMLN}{\ensuremath{\alpha^{{\rm ML}}_{{\rm N}}}}

\newcommand{\alphaMLt}{\ensuremath{\tilde{\alpha}^{{\rm ML}}}}

\newcommand{\alphaHN}{\ensuremath{\alpha^{{\rm H}}_{{\rm N}}}}
\newcommand{\betaHN}{\ensuremath{\beta^{{\rm H}}_{{\rm N}}}}
\newcommand{\varphilt}{\ensuremath{\varphi^{{\rm lt}}}}
\newcommand{\varphiL}{\ensuremath{\varphi^{{\rm L}}}}
\newcommand{\varphiLlt}{\ensuremath{\varphi^{{\rm L, lt}}}}
\newcommand{\varphiHlt}{\ensuremath{\varphi^{{\rm H, lt}}}}
\newcommand{\varphiH}{\ensuremath{\varphi^{{\rm H}}}}
\newcommand{\piL}{\ensuremath{\pi^{{\rm L}}}}
\newcommand{\piH}{\ensuremath{\pi^{{\rm H}}}}
\newcommand{\pilt}{\ensuremath{\pi^{{\rm lt}}}}
\newcommand{\piLlt}{\ensuremath{\pi^{{\rm L, lt}}}}
\newcommand{\piHlt}{\ensuremath{\pi^{{\rm H, lt}}}}
\newcommand{\PiL}{\ensuremath{\Pi^{{\rm L}}}}
\newcommand{\PiH}{\ensuremath{\Pi^{{\rm H}}}}
\newcommand{\PhiL}{\ensuremath{\Phi^{{\rm L}}}}
\newcommand{\PhiH}{\ensuremath{\Phi^{{\rm H}}}}

\newcommand{\NL}{\ensuremath{ N^{\rm L}}}
\newcommand{\NML}{\ensuremath{ N^{ {\rm ML}}}}
\newcommand{\NH}{\ensuremath{ N^{\rm H}}}

\title{Scalar two-point functions at the late-time boundary of de Sitter}
\author[a,b]{Gizem \c{S}eng\"or}
\author[a]{Constantinos Skordis}
\affiliation[a]{CEICO, Institute of Physics of the Czech Academy of Sciences,\\ Na Slovance 1999/2, 182 21 Praha 8, Czechia}
\affiliation[b]{Physics Department, Boğaziçi University\\
34342 Bebek, İstanbul Turkey}
\emailAdd{gizem.sengor@boun.edu.tr}
\emailAdd{skordis@fzu.cz}
\abstract{
We calculate two-point functions of scalar fields of mass $m$ and their conjugate momenta at the late-time boundary of de Sitter with Bunch-Davies boundary conditions, in general $d+1$ spacetime dimensions.
We perform the calculation using the wavefunction picture and using canonical quantization.
With the latter one clearly sees how the late-time field and conjugate momentum operators are linear combinations of the normalized 
late-time operators $\alphaN$ and $\betaN$ that correspond to unitary irreducible representations of the de Sitter group with well-defined inner products.
The two-point functions resulting from these two different methods are equal and 
we find that both the autocorrelations of $\alphaN$ and $\betaN$ and their cross correlations contribute to the late-time field and conjugate momentum two-point functions. This happens 
both for light scalars ($m<\frac{d}{2}H$), corresponding to complementary series representations, and heavy scalars ($m>\frac{d}{2}H$), corresponding to principal series representations of the de Sitter group,
where $H$ is the Hubble scale of de Sitter. In the special case $m=0$, only the $\betaN$ autocorrelation contributes to the conjugate momentum two-point function in any dimensions and we gather hints that suggest $\alpha_N$ to correspond to discrete series representations for this case at $d=3$.
}

\begin{document}
	\maketitle
	
	\newpage
	\section{Introduction}

	Within the framework of quantum field theory 
 we describe elementary particles propagating on Minkowski spacetime in terms of unitary irreducible representations of the Poincaré group \cite{10.2307/1968551}. 
The Poincaré group arises in this situation because it is the symmetry group of four dimensional Minkowski spacetime. 
There is another group, however, which is physically relevant, particularly in the case of cosmic acceleration pertaining 
 expanding spacetimes during the era of cosmic inflation in the early Universe or the era of dark energy domination in the late Universe.
This is the group $SO(d+1,1)$ which is isomorphic to the conformal group of $d-$dimensional Euclidean space and the group of isometries of $d+1$-dimensional de Sitter spacetime.
	
	Starting with the work of Harish-Chandra (see \cite{herb1991} for a pedagogical overview of his work and \cite{harish-chandra1955,borel1961,harish-chandra1963,harish-chandra1965,harish-chandra1966,harish-chandra1970} for the original articles), the study of the unitary irreducible representations of $SO(d+1,1)$ is a well established subject of investigation within representation theory. 
In \cite{Sengor:2019mbz},  following \cite{Dobrev:1977qv, Basile:2016aen} on the representation theory of $SO(d+1,1)$,
 we established a connection between the scalar unitary irreducible representations of $SO(d+1,1)$ and the late-time profile of scalar fields of mass $m$ on de Sitter. 
Letting $H$ be the Hubble scale of de Sitter, we focused on light scalar fields ($m<\frac{d}{2}H$) corresponding to complementary series representations and  heavy scalar fields ($m>\frac{d}{2}H$) 
 corresponding to principal series representations.
A distinguishing feature between these two representation classes is that the inner product is defined differently for each one. 
Furthermore, except from the special case lying between these two representation classes, that is $m=\frac{d}{2}H$~\footnote{The case  $m=\frac{d}{2}H$ 
is not captured by this current work.}, 
for both representation classes there exist two operators $\alpha$ and $\beta$, for each $m$. Having found these late-time operators, two questions arise: 
	\begin{enumerate}
	\item Can we find a relation between canonical field and momenta and the late-time operators?
	\item What are the correlation functions of these late-time operators?
	\end{enumerate}
Answering the first question can help in familiarizing one further with the nature of these de Sitter late-time boundary operators. Answering the second question 
can be insightful in constructing and exploring the nature of holographic duals to field theories on de Sitter in the long run.  
	
	Holography is an important tool in approaching quantum fields on curved spacetimes \cite{tHooft:1993dmi,Susskind_1995}. Its first realization came in the form of a correspondence between a gravitational theory on Anti de Sitter (AdS) and a Conformal Field Theory (CFT), using methods of string theory and supersymmetry \cite{Maldacena:1997re}. 
 The  AdS/CFT duality has found numerous applications.   A recent example is the application of AdS/CFT in the interpretation of gravitational wave observations \cite{demircik2020rapidly}, 
which discusses how the equation of state of dense cores of neutron stars derived by modelling the QCD phase at the core via AdS/CFT~\cite{Ishii:2019gta},
 can help in determining the nature of the unidentified compact object, tagged  GW190814,  recently observed by the LIGO and VIRGO collaborations~\cite{Abbott_2020}. 
	
	Looking at the merits of AdS/CFT, one begins to wonder about the nature of holography in more generality.
 The Anti de Sitter spacetime is one of the three maximally symmetric spacetimes whose symmetry group involves conformal symmetries. Approaching Holography from the perspective of symmetries brings on the question 
whether the other two maximally symmetric spacetimes, Minkowski and de Sitter, are also holographic. A Minkowski/CFT proposal was introduced in  \cite{de_Boer_2003}, by focusing on the 
symmetries near the null infinity. Around that time, the fact that the isometry group of de Sitter $SO(d+1,1)$ is also the conformal group of $d$-dimensional Euclidean space 
initiated the dS/CFT proposal \cite{Witten:2001kn,Strominger:2001pn}. Shortly after its introduction, this proposal was conjectured to be useful as a means to calculate 
inflationary spectra \cite{Maldacena:2002vr,vanderSchaar:2003sz}, in classifying inflationary potentials \cite{Binetruy:2014zya} and 
in interpretating the observed primordial spectral index and the tensor-to-scalar ratio of the cosmic microwave background radiation (CMB) 
angular power spectra~\cite{Planck:2018jri} as a hierarchy on the inflationary slow-roll parameters \cite{Pajer_2017}. 
These investigations signal that an improved understanding of de Sitter in the context of holography
 is bound to be instrumental in creating theoretical frameworks for a better description of our Universe. 
	
	There exists a single example that attempts to realize the dS/CFT proposal by considering Vasiliev Higher Spin Theory on de Sitter \cite{Anninos:2011ui}.  This first attempt proposes an $Sp(N)$ vector model as the dual theory. It has been discussed how the de Sitter spacetime can emerge from the degrees of freedom of the free $Sp(N)$ vector model under the flow equation approach in \cite{yokoyama2020holographic}. However, the $Sp(N)$ proposal has recently been revisited due to its lack of providing a well-defined Hilbert space and has been upgraded to a bosonic $O(2N)$ vector model \cite{Anninos:2017eib}. 
A key feature of the Vasiliev Higher Spin theory on de Sitter to note is that its scalar sector hosts only a single conformally coupled scalar field. Parallel to this discussion, there have been further promising developments on the dS/CFT literature in terms of a wavefunction formalism. The structure of the Bunch-Davies wavefunction for massless scalars on de Sitter have been considered in \cite{Anninos:2011kh} and its late-time structure for scalars of general mass and gauge fields in \cite{Anninos:2014lwa}. 
	
	By studying the properties of late-time operators for de Sitter guided by aspects of the de Sitter group, we hope to provide further insight on the intrinsic properties of quantum field theory on de Sitter and the properties of Euclidean CFT's that can be dual to de Sitter. The late-time operators we have obtained from the bulk solutions in \cite{Sengor:2019mbz} themselves are not the operators of some CFT, but they are the ones who would have the same properties as the dual CFT operators. Therefore by identifying their scaling dimensions and correlation functions we hope to provide data from which possible dual operators and the CFT's that host them can be recognized. 
We do not expect that a single CFT will accommodate all of the late-time operators for each value of $m$, but that different CFTs will correspond to different late-time operators and values of $m$.
	
	It is straightforward to compute late-time two-point functions from our late-time operators $\alphaN,~\betaN$ normalized with respect to the proper inner product and answer our second question. We use the convention where $\alphaN$ denotes the operator with the lower scaling dimension and $\betaN$ denotes the operator with the higher scaling dimension. To better identify the nature of these operators, that is, to answer our first question, 
our task is to compare the two-point functions $\langle\alphaN(\vec{k})\alphaN(\vec{k}')\rangle$, $\langle\alphaN(\vec{k})\betaN(\vec{k}')\rangle$, $\langle\betaN(\vec{k})\betaN(\vec{k}')\rangle$, 
which we calculate in section \ref{sec:two point functions from the late-time operators}, with the two-point functions for the late-time profile of canonical field  and its conjugate momenta, in $k$-space. There are a number of methods for computing cosmological correlation functions and we follow two different approaches. In section \ref{sec:two-point functions via the wavefunction} we make use of the wavefunction method whose role in cosmological perturbation theory has been recently growing \cite{Arkani-Hamed:2017fdk,Benincasa:2019vqr,Gorbenko:2019rza} and calculate two-point functions of the late-time field profile $\Phi_{\vec{k}}$ and late-time conjugate momenta $\Pi_{\vec{k}}$. In section \ref{sec:Two-point functions via canonical quantization} we employ canonical quantization and compute the two-point functions of canonically quantized field modes $\varphilt_{\vec{k}}(\eta)$ 
and momenta $\pilt_{\vec{k}}(\eta)$ at late-times.
	
A general approach in using the wavefunction method on de Sitter is to compute the Euclidean Anti de Sitter (EAdS) partition function and analytically continue the result to de Sitter. While the present work was being completed, there appeared the work of \cite{Isono2020} who also consider the de Sitter wavefunction for scalar fields. Similar to this work we carry all our calculations on de Sitter, in Poincaré coordinates where the metric takes the form
\be ds^2=\frac{-d\eta^2+d\vec{x}^2}{H^2|\eta|^2},\ee
with $\vec{x}\in\mathbb{R}^d$, $\eta\in(-\infty,0]$.
 Our section \ref{sec:Bunch Davies Wavefunction} can be considered as a short review of some of their results. In complementary series two-point functions one can neglect the contribution of the operator with the higher scaling dimension, $\betaN$, and the authors of \cite{Isono2020} do make this simplification. However, we are interested to track how both operators enter the two-point functions precisely and for this purpose we do not eliminate the contribution of the operator with the higher scaling dimension in the complementary series case. We refer the interested reader to \cite{Sleight:2021plv} for a careful treatment of analytic continuation from EAdS, especially when interactions are under consideration.
	
	Our main conclusion is that the operators $\alphaN(\vec{k})$ and $\betaN(\vec{k})$ in both case of the complementary and in the case of the principal series representations,
 contribute to the two-point functions calculated using the wavefunction or the canonical quantization method. 
The latter makes the relation of the field and conjugate momentum operators with the late-time operators explicit. In general, all of the two-point functions $\langle\alphaN\alphaN\rangle$, $\langle\betaN\betaN\rangle$, including the cross correlator $\langle\alphaN\betaN\rangle$, contribute both to the field and the conjugate momentum two-point function. Only in the case of a massless scalar  
the expressions simplify and the canonical momentum two-point function traces just the $\langle\beta^M_N\beta^M_N\rangle$. Based on these results, one may consider the late-time operator two-point functions to be the more elementary pieces than the field two-point functions. 

This paper is organized as follows: in section \ref{sec:Bunch Davies Wavefunction} we review the Bunch-Davies wavefunction on de Sitter and compute it for light scalars in \ref{subsec:The Bunch Davies wavefunction for light scalars} and for heavy scalars in \ref{subsec:The Bunch Davies wavefunction for heavy scalars}. We use this to
to compute the two-point functions  for light scalars in section \ref{subsubsec:The complementary series wavefunction} and for heavy scalars in section \ref{subsubsec:From the principal series wavefunction}. In section \ref{sec:Review of late-time boundary operators} we review the late-time boundary operators and normalize them with respect to the appropriate inner product in the case of light and heavy scalars individually. We enlarge our previous list of principal and complementary series late-time operators by pointing out how the massless scalar corresponds to \emph{exceptional series representations} in certain dimensions and how on $dS_4$ it hosts \emph{discrete series} late-time operators by discussing the exceptional series inner product and invariant subspaces involved in the cases of exceptional and discrete series in section \ref{subsec:Normalized operators corresponding to massless fields}.  We compute the two-point functions for light and heavy scalars individually in section \ref{sec:two point functions from the late-time operators}.
 The light scalars in the complementary series representation involve a shadow sector with operators of dimension $\tilde\Delta$ and a nonshadow sector with operators of dimension $\Delta$,
 where these two dimensions satisfy $\Delta+\tilde\Delta=d$, and the two operators are related by intertwining. We leave the further discussion of the shadow two-point functions to the appendix \ref{appendix:shadowtwopoint}. 
In section \ref{subsec:Review of canonical quantization} we review the canonical quantization picture and introduce field and conjugate momentum operators in section \ref{subsubsection:Canonically quantized field and momenta for light fields} for light scalars and in section \ref{subsubsection:Canonically quantized field and momenta for heavy fields} for heavy scalars. We calculate the two-point functions in two cases respectively in sections \ref{subsec:canonicalquantlatetimetwopoint_light} and \ref{subsec:canonicalquantlatetimetwopoint_heavy}. 
While in our main discussion we implement the late-time limit from the start, in appendix \ref{app:latetime at the end} we confirm that doing the calculations in the bulk and taking the late-time limit at the end matches the results with taking the late-time limit from the beginning. The two-point functions obtained via these three methods are compared in section \ref{sec:conclusions} where we summarize our conclusions and motivate future work. 

\section{Two-point functions via the wavefunction}
\label{sec:two-point functions via the wavefunction}
\subsection{Review of the Bunch Davies Wavefunction for a scalar field on $dS_{d+1}$}
	\label{sec:Bunch Davies Wavefunction}
		The Bunch-Davies Wavefunction is a functional of a given late-time field profile, and is obtained by the path integral over all field configurations that behave regularly at early times and approach the given profile $\Phi$ at some late-time $\eta_0$
	\be \label{BDW} \Psi_{BD}[\Phi,\eta_0]=\int^{\phi(\eta_0)=\Phi}_{\substack{\phi(\eta)\to0\\
		\text{as}~\eta\to -\infty(1+i\epsilon)}} \mathcal{D}\phi e^{iS[\phi]}.\ee
		Here we neglect the backreaction of the scalar field to the geometry and take the metric to be frozen to de Sitter metric. The mode functions of the scalar field in $k$-space 
are related to Hankel functions of first kind and a small factor of $i\epsilon$ is included at early times to guarantee that the modes vanish in this limit. We will refer to these boundary conditions first defined in \cite{Bunch:1978yq}, as \emph{Bunch-Davies boundary conditions}. 
		
		The general solutions to the classical equations of motion are Bessel functions. Among these solutions, Hankel functions of first kind $H^{(1)}_{\nu}(k|\eta|)$ in the case of light scalars with $\nu^2=\frac{d^2}{4}-\frac{m^2}{H^2}$ and $\tilde{H}^{(1)}_\rho(k|\eta|)$ in the case of heavy scalars with $\rho=\frac{m^2}{H^2}-\frac{d^2}{4}$ both behave as $H^{(1)}_{\nu}(k|\eta|),\tilde{H}^{(1)}_\rho(k|\eta|)\sim e^{ik|\eta|}$ in the limit $\eta\to-\infty$. Since these are of oscillatory nature, we need to regularize the lower limit as $\eta\to-\infty(1\pm i\epsilon)$, where the choice of upper sign is appropriate for $H^{(1)}_\nu(k|\eta|), \tilde{H}^{(1)}_\rho(k|\eta|)$. Another choice would be to use Hankel functions of second kind $H^{(2)}_{\nu}(k|\eta|), \tilde{H}^{(2)}_\rho(k|\eta|)$ which behave as $H^{(2)}_\nu(k|\eta|), \tilde{H}^{(2)}_\rho(k|\eta|)\sim e^{-ik|\eta|}$ in the limit $\eta\to-\infty$. For this choice the appropriately regularized boundary condition is the one with the lower sign. In practice it is not possible to carry out the formal integration in equation \eqref{BDW}, and one consults to a semiclassical approximation \cite{PhysRevD.28.2960} 
	\be \Psi_{BD}[\Phi,\eta_0]\sim e^{iW[\Phi,\eta_0]} \ee
	where $W[\Phi,\eta_0]$ is the Hamilton-Jacobi functional evaluated on the classical solution to the equations of motion that satisfy the boundary conditions of \eqref{BDW}. The functional $W[\Phi,\eta_0]$ is actually the onshell action, where the classical equations of motion have been employed
	\be \label{HamiltonJacobi} W[\Phi,\eta_0]=S_{onshell}[\Phi,\eta_0].\ee
	In employing this relation one needs to pay attention to the choice of the time contour \cite{Isono2020}. The boundary condition we picked in \eqref{BDW} corresponds to the choice $\mathcal{C}_{-}:\eta=\left(-(1+i\epsilon)\infty,\eta_0\right]$ on the time contour; while the other boundary condition corresponds to the time contour $\mathcal{C}_{+}:\eta=\left(-(1-i\epsilon)\infty,\eta_0\right]$. For our choice of boundary conditions along the contour $\mathcal{C}_-$, the onshell action in general dimensions is
\be 
   \label{Sonshellintroduction}
S_{onshell}=-\frac{1}{2}\int\frac{d^dk}{(2\pi)^{d}}\int_{\mathcal{C}_{-}}d\eta\frac{\partial}{\partial \eta} \left[a(\eta)^{d-1}\phi'_{\vec{k}}(\eta)\phi_{-\vec{k}}(\eta)\right],
\ee
where $a(\eta)=\frac{1}{H|\eta|}$.
Here the overall minus sign accompanies the contour $\mathcal{C}_{-}$ and $\left(\phi_{\vec{k}}\right)^*=\phi_{-\vec{k}}$ due to the reality conditions on the field.	

	In four dimensions, in $k$-space, following the compact notation of \cite{Anninos:2011kh}, the expression for the onshell action along the contour $\mathcal{C}_-$ takes the form
	\be \label{onshellintro} S_{onshell}[\Phi,\eta_0]=-\frac{1}{2}\int \frac{d^3k}{(2\pi)^{3}}a^2(\eta_0)\frac{v'_k(\eta_0)}{v_k(\eta_0)}|\Phi_{\vec{k}}|^2.\ee
	Here the mode functions\footnote{The solutions to what becomes of the Bessel's equation when $\nu=i\rho$ and the argument is real, are $\tilde{J}_{\rho}(x)$ and $\tilde{Y}_\rho(x)$ where $\rho\in\mathbb{R},x\in [0,\infty)$. These solutions are related to the Bessel functions by $\tilde{J}_\rho(z)= \sech\left(\frac{\rho\pi}{2}\right) \Real\left[J_{i\rho}(z)\right]$ 
and $\tilde{Y}_\rho(z)= \sech\left(\frac{\rho\pi}{2}\right) \Real\left[Y_{i\rho}(z)\right]$ \cite{doi:10.1137/0521055}. Out of these functions $$\tilde{H}^{(1)}_\rho(x)\equiv\tilde{J}_\rho(x)+i\tilde{Y}_\rho(x)=e^{-\rho\pi/2}H^{(1)}_{i\rho}(x).$$} are such that \cite{Sengor:2019mbz} 
	\be \label{lightmode} v_k(\eta)=|\eta|^{3/2}H^{(1)}_\nu(k|\eta|)~\text{with}~\nu^2=\frac{9}{4}-\frac{m^2}{H^2}~\text{for light scalars},\ee
	and
	\be \label{heavymode} v_k(\eta)=|\eta|^{3/2}\tilde{H}^{(1)}_\rho(k|\eta|)~\text{with}~\rho^2=\frac{m^2}{H^2}-\frac{9}{4}~\text{for heavy scalars}.\ee
	Here we left out the normalization for the mode function $v_k(\eta)$ as it does not enter into the calculation. Expanding out the expression \eqref{onshellintro}, 
the onshell action for a free scalar of general mass, is
 		\be \label{Sonshellgeneralm} S_{onshell}=-\frac{1}{2}\int \frac{d^3k}{(2\pi)^{3}}a^2(\eta_0)\left[\frac{3}{2\eta_0}+\frac{{\mathcal{H}}'(k\eta_0)}{\mathcal{H}(k\eta_0)}\right]|\Phi_{\vec{k}}|^2\ee
 		where prime denotes differentiation with respect to $\eta$ and $\mathcal{H}(k|\eta|)=H^{(1)}_\nu(k|\eta|)$ or $\mathcal{H}(k|\eta|)=\tilde{H}^{(1)}_\rho(k|\eta|)$ 
for the cases of light and heavy scalars respectively.


	 Another way to evaluate the wavefunction is to solve the Wheeler-DeWitt equations it obeys. In the ADM formulation of general relativity coupled to some field, $d+1$ out of $\frac{d(d+1)}{2}$ Einstein equations correspond to constraint equations. One of these constraints is obtained by varying the action with respect to the lapse function, denoted here by $C_N$, and the remaining $d$ are obtained from variation with respect to the shift vector, $C_i$. These constraints capture the diffeomorphism invariance of the theory. The wave function is also subject to these constraints
	 \begin{align}\label{constraints}
	 C_N\Psi&=0\\
	 C_i\Psi&=0.
	 \end{align}
	 By assuming $\Psi=e^{iS/\hbar}$, and expanding \eqref{constraints} order by order in $\hbar$, at zeroth order one arrives at the Hamilton-Jacobi equations for $S$. Along these lines, \cite{Pimentel:2013gza} provides a study of how the constraints on the wavefunction imply consistency conditions for the n-point functions in the case of inflation. 
	 
	 In the next two subsections we will review the wavefunction of light and heavy scalars individually without going into EAdS and make use of these results in the rest of our discussion. In doing so we will employ the asymptotic expansion of the Hankel function of first kind $H^{(1)}_\nu(z)=J_\nu(z)+iY_\nu(z)$ and $\tilde{H}^{(1)}_\rho=\tilde{J}_\rho(k|\eta|)+i\tilde{Y}_\rho(k|\eta|)$. The technical subtlety is in the difference between the asymptotic expansion of these two functions. Let us mention in passing that in EAdS, the mode functions that satisfy the given boundary conditions are related to Bessel  $K_\nu$ functions. We point interested readers towards \cite{doi:10.1137/0521055} for the properties of these functions.

	\subsubsection{Complementary series wavefunction for light scalars}
	\label{subsec:The Bunch Davies wavefunction for light scalars}
	We work in de Sitter spacetime in $(d+1)$ dimensions, where the on-shell action is given by \eqref{Sonshellintroduction}. We set
\be\label{phikind}\phi_{\vec{k}}=\Phi_{\vec{k}}\frac{v_k(\eta)}{v_k(\eta_0)},~~~v_k(\eta)=|\eta|^{d/2}H^{(1)}_\nu(k|\eta|),~~\nu^2=\frac{d^2}{4}-\frac{m^2}{H^2},\ee
where the factor
\be\label{bulktoboundary} \frac{v_k(\eta)}{v_k(\eta_0)}\equiv \mathcal{B}(\eta;\eta_0)\ee
is referred to as the \emph{bulk-to-boundary propagator} \cite{Anninos:2014lwa}. The evaluation of the onshell action boils down to evaluating the following expression
\be S^{compl}_{onshell}=-\frac{1}{2H^{d-1}}\int \frac{d^dk}{(2\pi)^{d}}|\eta_0|^{-d}|\Phi_{\vec{k}}|^2\left\{-|\eta_0|\frac{d}{d|\eta|}\left[\frac{|\eta|^{d/2}H^{(1)}_\nu(k|\eta|)}{|\eta_0|^{d/2}H^{(1)}_\nu(k|\eta_0|)}\right]_{|\eta|=|\eta_0|}\right\}\ee
where the curly bracket involves the differentiation of the bulk-to-boundary propagator. We are interested in correlators at the late-time limit, that is the limit $|\eta_0|\to0$. Thus we will employ the asymptotic expansion of $H^{(1)}_\nu(z)$ for small argument. For real index\footnote{In the case of integer index $\nu=n$, the Bessel function of second kind is modified \cite{arfken}
and the limit is 
\begin{align}
    \lim_{z\to 0}H^{(1)}_n(z)=&\frac{1}{\Gamma(n+1)}\left(\frac{z}{2}\right)^n\left[1+\frac{2i}{\pi}ln\left(\frac{z}{2}\right)-\frac{i}{\pi}\left(\psi(n+1)-\gamma_E\right)\right]-\frac{i}{\pi}\left(\frac{z}{2}\right)^{-n}\sum^{n-1}_{k=0}\frac{\Gamma(n-k)}{k!}\left(\frac{z^2}{4}\right)^k
\end{align}
where $\psi(m)=\Gamma'(m)/\Gamma(m)$, is the digamma function with $\psi(1)=-\gamma_E$. In the case when $\nu$ is not an 
integer, the series expansion involves terms of order $z^{-\nu+2}$ coming from the series expansion \eqref{Series_Y} and \eqref{Series_Y_Mathematica} shown in appendix-\ref{app:latetime at the end}.
These $z^{-\nu+2}$ terms are subleading next to the $z^{-\nu}$ term in the late-time limit and therefore we neglect them from the beginning for our purposes.} $\nu\in\mathbb{R}$, this expansion has the form \cite{NIST:DLMF}~\footnote{Note 
that this expansion holds for $\nu>0$. In \cite{Sengor:2019mbz} we identified two branches of operators, branch-I for solutions with $\nu>0$ and branch-II for solutions with $\nu<0$, where the difference was due to the different asypmtotic behavior of Bessel functions of positive and negative order. However, when we considered these solutions and the corresponding late-time operators, in terms of the unitary irreducible representations of $SO(d+1,1)$ we found that the two branches are related by a shadow transformation. For this reason it is enough to consider only one branch, and from the point of view of this section to focus only on the asymptotic behaviour for $\nu>0$. In \cite{Sengor:2019mbz} we neglected a subdominant contribution to the positive exponent term because in that work our main focus was on the momentum dependence rather than the normalization. Throughout this work we will include that subdominant term because we are also interested in the normalization. This correction will imply a correction for the late-time operator $\beta$ in section \ref{sec:two-point functions via the late-time boundary operators}.} 
\be
\label{Hcompl}
\lim_{|\eta|\to0}H^{(1)}_\nu(k|\eta|) = 
\frac{1+i\cot(\pi\nu)}{\Gamma(\nu+1)}\left(\frac{k|\eta|}{2}\right)^\nu-\frac{i\Gamma(\nu)}{\pi}\left(\frac{k|\eta|}{2}\right)^{-\nu}.
\ee
With this, the curly bracket term becomes
\begin{align}
\label{complementarykernel}
&\left\{-|\eta_0|\left[\frac{d}{d|\eta|}\mathcal{B}(\eta,\eta_0)\right]_{|\eta|=|\eta_0|}\right\}_{|\eta_0|\to0}=-\frac{\frac{1+i\cot(\pi\nu)}{\Gamma(\nu+1)}\left(\frac{d}{2}+\nu\right)\left(\frac{k|\eta_0|}{2}\right)^{2\nu}-\frac{i\Gamma(\nu)}{\pi}\left(\frac{d}{2}-\nu\right)}{\frac{1+i\cot(\pi\nu)}{\Gamma(\nu+1)}\left(\frac{k|\eta_0|}{2}\right)^{2\nu}-\frac{i\Gamma(\nu)}{\pi}}.
\end{align}
where following \cite{Isono2020} we have eliminated the lower exponent in this expression. Since in the complementary series the exponent $2\nu$ is a real number and 
since terms that involve $|\eta_0|^{2\nu}$ are negligible as $\eta_0\to0$, \cite{Isono2020} expanded the denominator leading to a simpler expression in that case.
%
 Here we kept these $|\eta_0|^{2\nu}$ terms in our final expressions further below and comment on their negligibility later on.
Putting everything together gives the wavefunction
\begin{align}
\label{complwavefunc}
\Psi^{compl}_{BD}\left[\Phi;\eta_0\right] 
&=\Ncal_c\exp\left\{
\frac{|\eta_0|^{-d}}{2H^{d-1}}\int\frac{d^dk}{(2\pi)^{d}}
\left[
 i \frac{d}{2} 
 + i\nu \frac{\frac{1+i\cot(\pi\nu)}{\Gamma(\nu+1)}\left(\frac{k|\eta_0|}{2}\right)^{2\nu} +  i \frac{\Gamma(\nu)}{\pi}}{\frac{1+i\cot(\pi\nu)}{\Gamma(\nu+1)}\left(\frac{k|\eta_0|}{2}\right)^{2\nu}-\frac{i\Gamma(\nu)}{\pi}}
\right]
|\Phi_{\vec{k}}|^2
\right\}.
\end{align}
The wavefunction has a time dependent normalization $\Ncal_c = \Ncal_c(\eta_0)$  which does not effect the correlation functions \cite{Eboli:1988qi,Guven:1987bx}. Note that while the late-time profile $\Phi$ is real, the Fourier modes $\Phi_{\vec{k}}$ it is decomposed into as $\Phi=\int\frac{d^dk}{(2\pi)^d}\Phi_{\vec{k}}e^{i\vec{k}\cdot\vec{x}}$, are complex valued. The reality condition of the late-time profile implies that $(\Phi_{\vec{k}})^*=\Phi_{-\vec{k}}$. 
The same reality condition also holds for bulk Fourier modes $(\phi_{\vec{k}})^*=\phi_{-\vec{k}}$.

	\subsubsection{Principal series wavefunction for heavy scalars}
	\label{subsec:The Bunch Davies wavefunction for heavy scalars}
	Late-time boundary operators for heavy scalars on de Sitter fall under the principal series representations. For these the index is purely imaginary $\nu=i\rho$, $\rho\in\mathbb{R}$. The on-shell action in general dimensions is
	\be S^{{\rm Prin}}_{onshell}=-\frac{1}{2H^{d-1}}\int \frac{d^dk}{(2\pi)^{d}}|\eta_0|^{-d}|\Phi_{\vec{k}}|^2\left\{-|\eta_0|\frac{d}{d|\eta|}\left[\frac{|\eta|^{d/2}\tilde{H}^{(1)}_\rho(k|\eta|)}{|\eta_0|^{d/2}\tilde{H}^{(1)}_\rho(k|\eta_0|)}\right]_{|\eta|=|\eta_0|}\right\}.\ee
	The bulk-to-boundary propagator becomes
	\be \mathcal{B}(\eta;\eta_0)\equiv\frac{v_k(\eta)}{v_k(\eta_0)}=\frac{|\eta|^{d/2}\tilde{H}^{(1)}_\rho(k|\eta|)}{|\eta_0|^{d/2}\tilde{H}^{(1)}_\rho(k|\eta_0|)}.\ee
	The late-time limit for the Hankel function of interest in this case is\footnote{In \cite{Sengor:2019mbz} we made use of the identity \cite{NIST:DLMF}
	\be 
    \label{complexgamma} 
  \Gamma(1+i\rho)=\sqrt{\frac{\pi\rho}{\sinh(\pi\rho)}}e^{i\gamma_{\rho}},
\ee
	where $\gamma_\rho$ is real and defined complex numerical coefficients $c_\rho$, $d_\rho$
	\begin{subequations}
	    \begin{align}
	        c_\rho+d_\rho&=\sqrt{\frac{2}{\pi\rho}}\left[\sqrt{tanh\left(\frac{\pi\rho}{2}\right)}+\sqrt{\coth\left(\frac{\pi\rho}{2}\right)}\right]\frac{e^{-i\gamma_\rho}}{2},\\
c_\rho-d_\rho&=\sqrt{\frac{2}{\pi\rho}}\left[\sqrt{tanh\left(\frac{\pi\rho}{2}\right)}-\sqrt{\coth\left(\frac{\pi\rho}{2}\right)}\right]\frac{e^{i\gamma_\rho}}{2}.
	    \end{align}
	\end{subequations}
	Rather than $c_\rho$, $d_\rho$, we will make use of the identity \eqref{complexgamma}.}
	\be 
   \lim_{|\eta|\to0}\tilde{H}^{(1)}_\rho(k|\eta|)=\frac{1+\coth(\rho\pi)}{\Gamma(1+i\rho)}e^{-\pi\rho/2}\left(\frac{k|\eta|}{2}\right)^{i\rho}-i\frac{\Gamma(i\rho)}{\pi}e^{-\pi\rho/2}\left(\frac{k|\eta|}{2}\right)^{-i\rho}.
\label{Hprinc}
\ee
This expansion is the main difference between the light and heavy scalars. With $\rho$ real, the complex conjugation of the Gamma functions work as $(\Gamma(i\rho))^*=\Gamma(-i\rho)$ and $(\Gamma(1+i\rho))^*=\Gamma(1-i\rho)$. 

Following through the same steps as in subsection \ref{subsec:The Bunch Davies wavefunction for light scalars} and simplifying the coefficients as much as possible, we obtain
\begin{align}\label{principalkernel}
&\left\{-|\eta_0|\left[\frac{d}{d|\eta|}\mathcal{B}(\eta,\eta_0)\right]_{|\eta|=|\eta_0|}\right\}_{|\eta_0|\to0}=-\frac{\left(\frac{d}{2}-i\rho\right)-\left(\frac{d}{2}+i\rho\right)e^{-\rho\pi-2i\gamma_\rho}\left(\frac{k|\eta_0|}{2}\right)^{2i\rho}}{1-e^{-\rho\pi-2i\gamma_\rho}\left(\frac{k|\eta_0|}{2}\right)^{2i\rho}}.
\end{align}
so that the principal series wavefunction is
\begin{align}
\Psi^{princip}_{BD} 
&=
\label{principwavefunc}
\Ncal_p \exp\left\{\frac{|\eta_0|^{-d}}{2H^{d-1}}\int\frac{d^dk}{(2\pi)^d}
\left[
i \frac{d}{2} + \rho \frac{ 1 + e^{-\rho\pi-2i\gamma_\rho}\left(\frac{k|\eta_0|}{2}\right)^{2i\rho} }{ 1 - e^{-\rho\pi-2i\gamma_\rho}\left(\frac{k|\eta_0|}{2}\right)^{2i\rho}}
\right]
 |\Phi_{\vec{k}}|^2\right\},
\end{align}
Taking the modulus square of $\Psi$ leads to an expression which  has the same form as that of \cite{Isono2020}.

	\subsection{Two-point functions}
	\label{sec:two-point functions}
	Once the wavefunction is known, the n-point function is easy to calculate. In this section we will give the general pattern for the calculations to follow. We will study the results for light scalars, in otherwords the complementary series representations and heavy scalars, or the principal series representations, in separate sections. We are interested in both the two-point function of the canonical field and its conjugate momenta. 

The on-shell action is in general
\be\label{consideringonshellasasum} S_{onshell}\left[\Phi_{\vec{k}}\right]=\int\frac{d^dk}{(2\pi)^d}\mathcal{L}\left[\Phi_{\vec{k}}\right]\sim\sum_{\vec{k}\in\mathbb{R}^{d+}}\mathcal{L}\left[\Phi_{\vec{k}}\right].\ee
The Bunch-Davies wavefunction is essentially a collection of individual wavefunctions for each $\vec{k}$. Hence, we may write it as
\begin{align}
\nn|\Psi_{BD}\left[\Phi_{\vec{k}}\right]|^2&=\Psi_{BD}\left[\Phi_{\vec{k}}\right]\Psi^*_{BD}\left[\Phi_{\vec{k}}\right]\\
\nn&=e^{i\sum_{\vec{k}\in\mathbb{R}^{d+}}\mathcal{L}\left[\Phi_{\vec{k}}\right]}e^{-i\sum_{\vec{k}\in\mathbb{R}^{d+}}\mathcal{L}^*\left[\Phi_{\vec{k}}\right]}\\
&\sim\prod_{\vec{k}\in\mathbb{R}^{d+}}e^{i\mathcal{L}\left[\Phi_{\vec{k}}\right]}e^{-i\mathcal{L}^*\left[\Phi_{\vec{k}}\right]}.
\end{align}
As we already pointed out, due to the reality conditions of the field, $\Phi_{\vec{k}}$ and $\Phi_{-\vec{k}}$ are not independent modes. This is why only the contribution of $\vec{k}\in\mathbb{R}^{d+}$ appears above. This fact also works into the measure for n-point functions.

We will use the notation  $\left[\mathcal{D}\Phi_{\vec{k}}\right]=\prod_{\substack{\vec{k}\in\mathbb{R}^{d+}}}d\Phi_{\vec{k}}$, for the integration measure. Then, a general n-point function is obtained by
\begin{align} \label{npointfunct} \langle \nn\Ocal^{(1)}_{\vec{k}_1}\dots \Ocal^{(n)}_{\vec{k}_n}\rangle
&=\frac{1}{\Ncal(\eta_0)}\int_{\substack{\text{configurations}\\ \text{where}~\Phi^*_{\vec{k}}=\Phi_{-\vec{k}}}}\left[\mathcal{D}\Phi_{\vec{k}}\right]\Psi^*_{BD}\left[\Phi_{\vec{k}}\right]\Ocal^{(1)}_{\vec{k}_1}...\Ocal^{(n)}_{\vec{k}_n}\Psi_{BD}\left[\Phi_{\vec{k}}\right]\end{align}
where
\be\label{normalizationofpsi}\Ncal(\eta_0)=\int_{\substack{\text{configurations}\\ \text{where}~\Phi^*_{\vec{k}}=\Phi_{-\vec{k}}}}\left[\mathcal{D}\Phi_{\vec{k}}\right]|\Psi_{BD}|^2=\int_{\substack{\text{configurations}\\ \text{where}~\Phi^*_{\vec{k}}=\Phi_{-\vec{k}}}}\left[\mathcal{D}\Phi_{\vec{k}}\right]\prod_{\vec{k}\in\mathbb{R}^{d+}}e^{i\mathcal{L}\left[\Phi_{\vec{k}}\right]}e^{-i\mathcal{L}^*\left[\Phi_{\vec{k}}\right]}.\ee
To simplify the notation we will not explicitly write the reality conditions in our expressions from now on.

In calculating the Gaussian integrals we make use of the identity \cite{Zee:706825} 
\be \label{Zeegaussint} \langle X_iX_j\rangle= \frac{\int dX_1\dots \int dX_N e^{-\frac{1}{2}\vec{X}^T\cdot A\cdot \vec{X}}X_iX_j}{\int dX_1\dots \int dX_N e^{-\frac{1}{2}\vec{X}^T\cdot A\cdot \vec{X}}}=\left(A^{-1}\right)_{ij}.\ee

The general form of the wavefunction depends on the amplitude of coordinate momenta $k$ and time-slice, which we will denote as $\Pcal(k,|\eta_0|)$, and on the field profile $\Phi_{\vec{k}}$ as follows
\be
\label{Psiform} 
\Psi_{BD}\left[\Phi,\eta_0\right]=\Ncal(\eta_0)\exp\left[-\frac{1}{2}\int\frac{d^dk}{(2\pi)^d}\Pcal(k,|\eta_0|)\Phi_{\vec{k}}\Phi_{-\vec{k}}\right].
\ee
In fact our notation here matches the notation of \cite{Guven:1987bx}. The $\Phi_{\vec{k}}$ dependence of the wavefunction, which has Gaussian form, determines the general form of the two-point functions while momentum and time dependence of $\Pcal(k,|\eta_0|)$, which is complex valued, determines the late-time behaviour of the results.

The two-point function for field operators in $k$-space is evaluated from
\be \label{field2point} \langle \Phi_{\vec{k}}\Phi_{\vec{k}'}\rangle=\frac{1}{\Ncal}\int\left[\mathcal{D}\Phi_{\vec{k}''}\right]\Phi_{\vec{k}}\Phi_{\vec{k}'}e^{-\frac{1}{2}\int\frac{d^dk''}{(2\pi)^d}\Bigg(\Pcal^*(k'',|\eta_0|)+\Pcal(k'',|\eta_0|)\Bigg)\Phi_{\vec{k}''}\Phi_{-\vec{k}''}}.\ee
Observe that we can write the argument of the exponent as 
\be  \frac{1}{(2\pi)^d}\left(\Pcal^*+\Pcal\right)\Phi_{\vec{k}''}\Phi_{-\vec{k}''}=\int d^dq \Phi_{\vec{k}''}\mathcal{A}(\vec{k}'',\vec{q}'')\Phi_{\vec{q}''}\ee
where 
\be \mathcal{A}(\vec{k}'',\vec{q}'')=\frac{\Pcal^*+\Pcal}{(2\pi)^d}\delta^{(d)}(\vec{k}''+\vec{q}''),\ee
\be \mathcal{A}^{-1}(\vec{k}'',\vec{q}'')=\frac{(2\pi)^d}{\Pcal^*+\Pcal}\delta^{(d)}(\vec{k}''+\vec{q}'').\ee
This puts expression \eqref{field2point} in the form of \eqref{Zeegaussint} and we obtain
\begin{align} 
\label{field2pointgaussian}
\langle \Phi_{\vec{k}}\Phi_{\vec{k}'}\rangle=\mathcal{A}^{-1}(\vec{k},\vec{k}')=\frac{(2\pi)^d}{\Pcal^*(k)+\Pcal(k)}\delta^{(d)}(\vec{k}+\vec{k}').
\end{align}

In position space, the canonical momentum operator is identified as $\Pi=\frac{1}{i}\frac{\delta}{\delta\Phi}$. The commutation relation is evaluated by considering the action of canonical filed and momentum on any generic wavefunction, which gives
\begin{align}\label{PiPhicomm} \left[\Phi(\vec{x},\eta),\Pi(\vec{y},\eta)\right]=i\delta^{(d)}(\vec{x}-\vec{y}).\end{align}

The conjugate momentum operator in $k$-space is \cite{Eboli:1988qi} 
\be\label{conjmomk} \Pi_{\vec{k}}=\frac{1}{i}(2\pi)^d\frac{\delta}{\delta\Phi_{-\vec{k}}}.\ee
and satisfies the following commutation relation \be\label{PikPhikcomm}\left[\Phi_{\vec{k}},\Pi_{\vec{k}'}\right]=i(2\pi)^d\delta^{(d)}(\vec{k}+\vec{k}').\ee

The two-point function of conjugate momentum operators are evaluated from
\begin{align}
\langle\Pi_{\vec{k}}\Pi_{\vec{k}'}\rangle=\frac{1}{\Ncal}\int \left[\mathcal{D}\Phi_{\vec{k}''}\right]\Psi^*_{BD}\left[\Phi_{\vec{k}''}\right]\left(\frac{(2\pi)^d}{i}\frac{\delta}{\delta\Phi_{-\vec{k}}}\right)\left(\frac{(2\pi)^d}{i}\frac{\delta}{\delta\Phi_{-\vec{k}'}}\right)\Psi_{BD}\left[\Phi_{\vec{k}''}\right],
\end{align}
For expressions in position space we refer the interested reader to \cite{Jackiw:1988sf}. We have not written dependence on $|\eta_0|$ explicitly so as to keep the notation more focused on the variables of interest. The differentiation of the wavefunction gives
\be \label{diffwave} \frac{\delta\Psi_{BD}\left[\Phi_{\vec{k}''}\right]}{\delta\Phi_{-\vec{k}'}}=-\Psi_{BD}\left[\Phi_{\vec{k}''}\right]\frac{\Pcal(k',|\eta_0|)}{(2\pi)^d}\Phi_{\vec{k}'},\ee
\be\label{diffdiffwave}
\frac{\delta}{\delta\Phi_{-\vec{k}}}\frac{\delta\Psi_{BD}\left[\Phi_{\vec{k}''}\right]}{\delta\Phi_{-\vec{k}'}}=\Psi_{BD}\left[\Phi_{\vec{k}''}\right]\frac{\Pcal(k)}{(2\pi)^d}\frac{\Pcal(k')}{(2\pi)^d}\Phi_{\vec{k}}\Phi_{\vec{k}'}-\Psi_{BD}\left[\Phi_{\vec{k}''}\right]\frac{\Pcal(k')}{(2\pi)^d}\delta^{(d)}(\vec{k}+\vec{k}'),\ee
leading
 to the following result for the two-point function of conjugate momentum operator
\begin{align}\label{mommomk}
\nn \langle\Pi_{\vec{k}}\Pi_{\vec{k}'}\rangle&=-\Pcal(k)\Pcal(k')\langle\Phi_{\vec{k}}\Phi_{\vec{k}'}\rangle+(2\pi)^d\Pcal(k')\delta^{(d)}(\vec{k}+\vec{k}')\\
&=(2\pi)^d\frac{\Pcal^*(k)\Pcal(k)}{\Pcal^*(k)+\Pcal(k)}\delta^{(d)}(\vec{k}+\vec{k}').
\end{align}

\subsubsection{Two-point functions for light scalars}
\label{subsubsec:The complementary series wavefunction}
Using \eqref{Psiform} along with \eqref{complwavefunc} we read-off  $\Pcal(k)$ in the case of the complementary series as  
\be 
\Pcal^{{(\rm Comp)}}(k)=\frac{1}{|\eta_0|^dH^{d-1}}\frac{\frac{i\left(1+i \cot(\nu\pi)\right)}{\Gamma(\nu+1)}\left(\frac{d}{2}+\nu\right)\left(\frac{k|\eta_0|}{2}\right)^{2\nu}+\frac{\Gamma(\nu)}{\pi}\left(\frac{d}{2}-\nu\right)}{\frac{i\Gamma(\nu)}{\pi}-\frac{1+i\cot(\nu\pi)}{\Gamma(\nu+1)}\left(\frac{k|\eta_0|}{2}\right)^{2\nu}},
\ee
so that using \eqref{field2pointgaussian} and \eqref{mommomk} it leads to the following results for the late-time two-point functions
\begin{align}
\nn\langle\Phi^L_{\vec{k}}\Phi^L_{\vec{k}'}\rangle&=\frac{\pi}{4}(2\pi|\eta_0|)^dH^{d-1}\delta^{(d)}(\vec{k}+\vec{k}')\times\\
\label{2pointlightPhi}&~~~~~~~~~\times\left[\frac{1+\cot^2(\nu\pi)}{\Gamma^2(\nu+1)}\left(\frac{k|\eta_0|}{2}\right)^{2\nu}+\frac{\Gamma^2(\nu)}{\pi^2}\left(\frac{k|\eta_0|}{2}\right)^{-2\nu}-\frac{2\cot(\nu\pi)}{\nu\pi}\right]\\
\nn\langle\Pi^L_{\vec{k}}\Pi^L_{\vec{k}'}\rangle&=\frac{\pi}{4}\frac{(2\pi)^d\delta^{(d)}(\vec{k}+\vec{k}')}{|\eta_0|^dH^{d-1}}\times\Bigg[-\frac{2\cot(\nu\pi)}{\nu\pi}\left(\frac{d}{2}+\nu\right)\left(\frac{d}{2}-\nu\right)\\
\label{compl2ptPi}&+\frac{1+\cot^2(\nu\pi)}{\Gamma^2(\nu+1)}\left(\frac{d}{2}+\nu\right)^2\left(\frac{k|\eta_0|}{2}\right)^{2\nu}+\frac{\Gamma^2(\nu)}{\pi^2}\left(\frac{d}{2}-\nu\right)^2\left(\frac{k|\eta_0|}{2}\right)^{-2\nu}\Bigg].
\end{align}
We discuss the case where $\nu$ is an integer in the appedix \ref{app:latetime at the end}.

\subsubsection{Two-point functions for heavy scalars}
\label{subsubsec:From the principal series wavefunction}
Using \eqref{Psiform} along with \eqref{principwavefunc} we read-off  $\Pcal(k)$ in the case of the Principal series as  
\be
 \label{Pprinc}
\Pcal^{({\rm Prin})}(k)=-\frac{|\eta_0|^{-d}}{H^{d-1}}\frac{\left(i\frac{d}{2}+\rho\right)-\left(i\frac{d}{2}-\rho\right)e^{-\rho\pi-2i\gamma_\rho}\left(\frac{k''|\eta_0|}{2}\right)^{2i\rho}}{1-e^{-\rho\pi-2i\gamma_\rho}\left(\frac{k''|\eta_0|}{2}\right)^{2i\rho}},
\ee
leading to the following two-point functions
\begin{align}
\label{2pointheavyPhi}
\langle\PhiH_{\vec{k}}\PhiH_{\vec{k}'}\rangle
=&
 \frac{ (2\pi|\eta_0|)^dH^{d-1}}{4\rho \sinh(\rho\pi)} 
\left[2\cosh(\rho\pi)-e^{2i\gamma_\rho}\left(\frac{k|\eta_0|}{2}\right)^{-2i\rho}-e^{-2i\gamma_\rho}\left(\frac{k|\eta_0|}{2}\right)^{2i\rho}\right]
\nonumber
\\
& \ \ \ \   \times
\delta^{(d)}(\vec{k}+\vec{k}')
\\
\nn
\langle\PiH_{\vec{k}}\PiH_{\vec{k}'}\rangle =& \frac{(2\pi)^d\delta^{(d)}(\vec{k}+\vec{k}')}{|\eta_0|^dH^{d-1}}\frac{1}{4\rho \sinh(\rho\pi)}\times\Bigg[
 2 \left(\frac{d^2}{4}+\rho^2\right)\cosh(\rho\pi)
\\
\label{princp2ptPi}
& \ \ \ \ -\left(\frac{d}{2}-i\rho\right)^2e^{2i\gamma_\rho}\left(\frac{k|\eta_0|}{2}\right)^{-2i\rho}-\left(\frac{d}{2}+i\rho\right)^2e^{-2i\gamma_\rho}\left(\frac{k|\eta_0|}{2}\right)^{2i\rho}\Bigg].
\end{align}	
where we have once more used \eqref{field2pointgaussian} and \eqref{mommomk}.

\section{Two-point functions via the late-time boundary operators}
\label{sec:two-point functions via the late-time boundary operators}	
\subsection{Review of late-time operators}
\label{sec:Review of late-time boundary operators}
In \cite{Sengor:2019mbz} where we identified operators at the late-time boundary of de Sitter for scalar fields of various mass, it was sufficient to consider $H=1$ for our purposes. We will continue to follow this convention here and fix the normalization of the operators such that their norm is one. While in \cite{Sengor:2019mbz} we focused on principal and complementary series among the unitary irreducible representations, here we will also address discrete series operators. The crucial property to take into consideration is that the properly defined inner product for light scalars involves shadow transformation.\footnote{In our discussion we are considering representations built from finite group elements. On a complementary route, one can also work with the algebra in constructing the representations. \cite{10.2307/2415039} is a pedagogical example to the later method. In both cases one will be faced with the difference in the normalization between states of principal, complementary and discrete series representations. Another reference that explicitly discusses the nontrivial inner product for the complementary and discrete series case is \cite{2007}.} Shadow transformation is a similarity transformation that changes the scaling dimension of an operator. For an operator $\alpha(\vec{x})$ with scaling dimension $\Delta=\frac{d}{2}-\nu$, we will denote its shadow by a tilde $\tilde{\alpha}$ which in position space will have the scaling dimension $\tilde{\Delta}=\frac{d}{2}+\nu$. For complementary series, the shadow transformation leaves the trace invariant, and the operators $\alpha$ and $\tilde{\alpha}$ are equivalent to each other. The situation is more involved in the case of discrete series \cite{Dobrev:1977qv}, which we will not discuss in detail here. For further details and the range of definition for the operator that realizes this transformation, we refer the readers to \cite{Sengor:2019mbz} and references there in. Here we will only state the necessary results whenever the shadow operators are involved. Thus we will work out the normalization individually for light and heavy scalars. 

The late-time operators are identified from the late-time limit of the field based on the following relation
  \begin{align}
      \lim_{|\eta_0|\to0}\phi(\vec{x},\eta)=\int\frac{d^dk}{(2\pi)^d}\left[|\eta_0|^{\frac{d}{2}-\mu}\alpha(\vec{k})+|\eta_0|^{\frac{d}{2}+\mu}\beta(\vec{k})\right]e^{i\vec{k}\cdot\vec{x}}
  \end{align}
  where in the case of light scalars $\mu=\nu$ and in the case of heavy scalars $\mu=i\rho$.

\subsubsection{Normalized operators corresponding to light scalars}
 \label{subsec:Normalized operators corresponding to light fields}
 
  In \cite{Sengor:2019mbz} we identified two branches of operators for light scalars $\{\alpha^{{\rm I}}(\vec{k}),\beta^{{\rm I}}(\vec{k})\}$
 and $\{\alpha^{{\rm II}}(\vec{k}),\beta^{{\rm II}}(\vec{k})\}$, however, these branches were related to each other via shadow transformations.
 Here we will only focus on Branch-I and in place of the label I, we will introduce a superscript label to distinguish the operators that correspond to light scalars in this section from the operators that will correspond to heavy scalars next. With overall coefficients related to mass of the field, these operators are\footnote{Here and in the next section $N_{\alpha}$ and $N_{\beta}$ denote the overall normalization, which we will shortly fix below, the operators $a_{\vec{k}}$ and $a^\dagger_{\vec{k}'}$ are respectively the annihilation and creation operators that satisfy the commutation relation $\left[a_{\vec{k}},a^\dagger_{\vec{k}'}\right]=(2\pi)^d\delta^{(d)}(\vec{k}-\vec{k}')$, $\nu^2=\frac{d^2}{4}-m^2$ with mass understood to be in terms of $H$ and $\nu$ denotes the positive root.}
 \begin{align}
 \alphaL(\vec{k})&=-\frac{i}{\pi}\Gamma(\nu) \NL_\alpha
 \left[a_{\vec{k}}-a^\dagger_{-\vec{k}}\right]\left(\frac{k}{2}\right)^{-\nu}\\
 \betaL(\vec{k})&=\frac{\NL_\beta}{\Gamma(\nu+1)}\left[\left(1+i\cot(\pi\nu)\right)a_{\vec{k}}+\left(1-i\cot(\pi\nu)\right)a^\dagger_{-\vec{k}}\right]\left(\frac{k}{2}\right)^\nu.
 \end{align}
 These operators correspond to light scalars $0<m^2<\frac{d^2}{4}$, which belong to the complementary series representations of de Sitter group $SO(d+1,1)$.
 The appropriately defined inner products are 
 \be\label{normalization}
\left(\alphaL(\vec{k}),\alphaLt(\vec{k})\right)=\frac{1}{\Omega}\int \frac{d^dk}{(2\pi)^{d}}\langle\alphaL(\vec{k})|\alphaLt{(\vec{k})}\rangle
\ee
where $\Omega\equiv\int\frac{d^dk}{(2\pi)^{d}}\langle -k|-k\rangle$, and the same expression also holds for $\betaL$. Without the shadow operator $\tilde{\alpha}$ the integrand in \eqref{normalization} is divergent. With the well-defined inner product the states obtained from the late-time operators are normalizable up to a dirac delta function whose effect we take into account in \eqref{normalization} by the volume factor $\Omega$.\footnote{Our notation for $\left(\mathcal{O}^L(\vec{k}),\tilde{\mathcal{O}}^L(\vec{k})\right)$ in \cite{Sengor:2019mbz} exactly matched the notation of \cite{Dobrev:1977qv}. Here we adjust our notation so as to take into account the factor of $\Omega$ for the normalization.} Including a brief summary in appendix \ref{appendix:shadowtwopoint} while referring the interested reader to \cite{Sengor:2019mbz} for further details of the shadow transformation and noting that $|0\rangle$ is the vacuum state annihilated by $a_{\vec{k}}$, the states involved in the inner product are
 \begin{align}
 |\alphaL({\vec{k}})\rangle\equiv \alphaL(\vec{k})|0\rangle&=\frac{i\Gamma(\nu)}{\pi} \NL_\alpha\left(\frac{k}{2}\right)^{-\nu}|-\vec{k}\rangle,\\
 |\alphaLt({\vec{k}})\rangle\equiv\alphaLt(\vec{k})|0\rangle&=\frac{i\Gamma(\nu)}{\pi} \NL_\alpha\left(k\right)^\nu|-\vec{k}\rangle.
 \end{align} 
 Demanding these states be normalized by \eqref{normalization}, sets $\NL_\alpha=\frac{\pi}{2^{\nu/2}\Gamma(\nu)}$. This also washes away any $H$ dependence expected from a dimensional analysis. Following the same argument for $\betaL$, $|\NL_{\beta}|^2=\frac{2^{\nu}\Gamma^2(\nu+1)}{1+\cot^2(\pi\nu)}$.We will set $\NL_{\beta}=\frac{2^{\nu/2}\Gamma(\nu+1)}{1-i\cot(\pi\nu)}$ and  the properly normalized late-time boundary operators in the complementary series become
\begin{subequations}
\label{normalizedcomp}
\begin{align}
\alphaLN(\vec{k})&=-i2^{\nu/2}\left[a_{\vec{k}}-a^\dagger_{-\vec{k}}\right]k^{-\nu}
\\
\betaLN(\vec{k})&=2^{-\nu/2}\left[\frac{1+i\cot(\pi\nu)}{1-i\cot(\pi\nu)}a_{\vec{k}}+a^\dagger_{-\vec{k}}\right]k^{\nu}.
\end{align}
\end{subequations}
Before moving on, let us also look at the commutation relation between these operators. The only nonzero commutation relation is
\be\label{complcomm} \left[\betaLN(\vec{k}),\alphaLN(\vec{k}')\right]=\frac{2i}{1-i\cot(\nu\pi)}(2\pi)^d\delta^{(d)}(\vec{k}+\vec{k}').\ee
This resembles the commutation relation between canonical momenta and canonical field operator.

\subsubsection{Normalized operators corresponding to massless scalars}
\label{subsec:Normalized operators corresponding to massless fields}

In the cases $c=\pm\frac{d}{2}$ the normalization of the intertwining operators $G_\chi$ and $G_{\tilde{\chi}}$, used in complementary series case, reviewed in section \ref{appendix:shadowtwopoint}, becomes problematic. The majority of these representations belong to what are called the \emph{exceptional series representations} which involve intertwining operators different than the complementary series intertwining operators. This happens whenever the following choice of normalization
\be
 n_+(\chi)=n_+(l,c)=\left(\frac{d}{2}+l+c-1\right)\frac{\Gamma(\frac{d}{2}+c-1)}{\Gamma(-c)},
\ee
has poles. The Gamma function $\Gamma(z)$ has poles at $z=0,-1,-2,-3,...$

Here we will mention details of the exceptional series following \cite{Dobrev:1977qv} only to the extend of the results we will make use of. We refer interested readers to \cite{sun2021note, Letsios:2023awz} for more detailed summaries which involve exceptional and discrete series for non zero spin. In general exceptional series representations are reducible \cite{Dobrev:1977qv}. In two and four spacetime dimensions, exceptional series representations of symmetric traceless tensors correspond to the unitary irreducible \emph{discrete series representations} \cite{sun2021note}. A very brief review of discrete series representations can be found in \cite{Sengor:2022lyv}. Our goal for this subsection is to identify late-time operators of a massless scalar on $dS_4$ as an example to discrete series representations. But we start our discussion in general dimensions with exceptional series case.

There are four categories of exceptional series representations labeled by $l$, for the compact subgroup $M=SO(d)$ label with values $l=0,1,...$, and $\sigma=1,2,...$ which work into the scaling weight $c$. The corresponding function spaces and their invariant subspaces labeled by $\chi=\{spin,c\}$ have the following $l$ and $\sigma$ dependence as listed in table \ref{exceptional table}

\vspace*{0.3cm}
\begin{table}[h]
\begin{center}
\begin{tabular}{|c|c|c|}
\hline
Exceptional series category & function space & invariant subspace\\
\hline
$\chi^-_{l_\sigma}=\{l,1-\frac{d}{2}-l-\sigma\}$ & $C^-_{l\sigma}$ & $E_{l\sigma}$\\
\hline
$\chi^+_{l_\sigma}=\{l,\frac{d}{2}+l+\sigma-1\}$ & $C^+_{l\sigma}$ & $F_{l\sigma}$\\
\hline
$\chi^{'-}_{l_\sigma}=\{l+\sigma,1-\frac{d}{2}-l\}$ & $C^{'-}_{l\sigma}$ & $F'_{l\sigma}$\\
\hline
$\chi^{'+}_{l_\sigma}=\{l+\sigma,\frac{d}{2}+l-1\}$ & $C^{'+}_{l\sigma}$ & $D_{l\sigma}$\\
\hline	
	\end{tabular}
 \end{center}
\caption{\label{exceptional table}Function spaces hosting the exceptional series representations.}
 \end{table}
\vspace*{0.3cm}
Notice that for the categories $\chi^\pm_{l_\sigma}$ spin matches the $SO(d)$ label, while it becomes more intricate for the categories $\chi^{'\pm}_{l_\sigma}$. 

For $d>2$ the invariant subspaces are nontrivial and these are reducible representations. Moreover the representations in one category can be obtained from the other by a chain of intertwining maps, also known as shadow transformations. These transformations involve mirror images. In some cases mirror images of the representations are not equivalent or are only partially equivalent. The definitions of the intertwining operators for inner products of these representations involve different subtleties in each case.

One of the late-time operators that correspond to a massless scalar is
\be \label{masslessalpha} 
 \alphaML(\vec{k})=-\frac{i}{\pi}\Gamma\left(\frac{d}{2}\right) \NML_\alpha
 \left[a_{\vec{k}}-a^\dagger_{-\vec{k}}\right]\left(\frac{k}{2}\right)^{-\frac{d}{2}},~~\text{with}~~c=-\frac{d}{2}.\ee
With $\chi=\{l,c\}$, and considering that $\sigma=1,2,..$ this can only belong to the $C^-_{01}$ function space of the $\chi^{-}_{0_1}$ category with $\sigma=1$ which has the invariant subspace $E_{01}$. Thus while we identified the rest of $\alphaL$ to belong to the complementary series, we are identifying $\alphaML$ in the case of the massless field to belong to the exceptional series. This identification is in agreement with what is also explicitly stated on page 99 of \cite{Dobrev:1977qv}. Moreover, in this case the normalization of the complementary series intertwining operator becomes
\be n_+\left(0,-\frac{d}{2}\right)=-\frac{\Gamma(-1)}{\Gamma(\frac{d}{2})},\ee
which is ill defined in any dimension because $\Gamma(-1)$ has a pole.

There is a well defined intertwining operator acting on $C^-_{l_\sigma}$ (as shown in chapter II, section 6.B, Figure 1 in \cite{Dobrev:1977qv})
\be G^+_{\chi^+_{l\sigma}}:C^-_{l\sigma}\to C^{+}_{l\sigma},\ee with an explicit expression in momentum space \cite{Dobrev:1977qv}
\be G^+_{\chi^+_{l\sigma}}(k)=\left(\frac{k^2}{2}\right)^{l+\sigma+\frac{d}{2}-1}\sum_{s=0}^{l}\frac{(d+l+s+\sigma-3)!(\sigma+l-s)!}{(d+2l+\sigma-3)!\sigma!}(-1)^{l-s}\Pi^{ls}(k)\ee
where $\Pi^{ls}(k)$ are projection operators and $\Pi^{00}(k)=1$. For our case of interest
\be G^+_{\chi^+_{01}}(k)=\left(\frac{k^2}{2}\right)^{\frac{d}{2}}. \ee
As this intertwining operator can act on $\alphaML(\vec{k})$ we have that
\begin{align}
\alphaMLt(\vec{k})&=G^+_{\chi^+_{01}}(k)\alphaML(\vec{k})\\
&=-\frac{i}{\pi}\Gamma\left(\frac{d}{2}\right) \NML_\alpha
\left[a_{\vec{k}}-a^\dagger_{-\vec{k}}\right]\left(k\right)^{\frac{d}{2}}.
\end{align}
Note that this intertwining operator establishes only partial equivalence.

Requiring
\begin{align}
\left(\alphaML,\alphaMLt\right)=\frac{1}{\Omega}\int \frac{d^dk}{(2\pi)^d}\langle\alphaML(\vec{k})|\alphaMLt(\vec{k})\rangle=1
\end{align} 
we read off the normalization to be $| \NML_\alpha|^2=\frac{\pi^2}{2^{d/2}\Gamma^2\left(\frac{d}{2}\right)}$. This is also the normalization factor we obtained for the rest of the light fields among complementary series $\alphaL$. Thus we obtain the following normalized operator 
\be
\alphaMLN(\vec{k})=-i2^{d/4}\left[a_{\vec{k}}-a^\dagger_{-\vec{k}}\right]k^{-d/2}.
 \label{alphaMN}
\ee 

The second of the late-time operators that correspond to a massless scalar is 
 \be \label{masslessbeta}
 \betaML(\vec{k})=\frac{\NML_\beta}{\Gamma(\frac{d}{2}+1)}\left\{ 
 \left[1+i\cot\left(\frac{\pi d}{2}\right)\right]a_{\vec{k}}+\left[1-i\cot\left(\frac{\pi d}{2}\right)\right]a^\dagger_{-\vec{k}}
\right\}\left(\frac{k}{2}\right)^{\frac{d}{2}}.
\ee
Only the representations of the category $\chi^+_{l_\sigma}$ can host this operator with $\sigma=1$. Therefore we identify $\betaML$ to belong to $C^+_{01}$. 
The function space $C^+_{l\sigma}$ has the invariant subspace $F_{l\sigma}$. The representations in $F_{0\sigma}$ are categorized to belong to the exceptional series for $d\geq 5$ \cite{Dobrev:1977qv}. In this case the well defined intertwining operators have workable explicit forms in position space. The normalization of the complementary series intertwining operator for $\betaML$ is
\be n_+\left(0,\frac{d}{2}\right)=(d-1)\frac{\Gamma(d-1)}{\Gamma(-\frac{d}{2})}.\ee
This normalization has poles for dimensions where $d$ is even and when $d=1$ it vanishes. Over all, which of the $\betaML$s belong to exceptional series depends on how many dimensions one is considering. In the case of $d=3$, relevant to inflationary cosmology, the $n_+$ normalization is \be n_+\left(0,\frac{3}{2}\right)=\frac{3}{2\sqrt{\pi}},\ee
and it seems that we can normalize $\beta^{{\rm ML}, d=3}$ with the complementary series intertwining operator. Therefore, normalized as in \ref{subsec:Normalized operators corresponding to light fields}
\be\label{normalizedBetaMasless} \beta^{ {\rm ML},d=3}_N(\vec{k})=2^{-\frac{3}{4}}\left[a_{\vec{k}}+a^\dagger_{-\vec{k}}\right]k^{\frac{3}{2}}.\ee

As we stated in the beginning of this subsection, in four spacetime dimensions exceptional series representations are the same as discrete series representations \cite{sun2021note}. Therefore with the above normalization properties and dimensions $\Delta=0,3$ respectively, we categorize $\alpha^{{\rm ML},d=3}_N$ 
to belong to discrete series representations while our analysis so far does not fully determine whether $\beta^{{\rm ML},d=3}_N$ belongs to 
complementary or discrete series. In principle the discrete series representations are identified as two disjoint sets $D^\pm_{l_\sigma}$ that the invariant subspace $D_{l_\sigma}$ of the category $\chi^{'+}_{l_\sigma}$ splits into \cite{Dobrev:1977qv}. But because we cannot fit spin-0 into $\chi^{'+}_{l_\sigma}$, we have to live with the invariant subspace $E_{0_1}$ of $\chi^{-}_{l_\sigma}$. Besides the difference in the inner product, discrete series representations have further properties, such as inequivalance of their shadows and the existence of a tower of highest weight and an independent tower of lowest weight states, 
that set them apart from the other categories. We leave it for future work to check whether our operators $\beta^{{\rm ML},d=3}_N(\vec{k}),\alpha^{{\rm ML},d=3}_N(\vec{k})$ 
satisfy these properties which can help identify both of them better. 

Note that the nontrivial commutation
\be 
\left[\beta^{{\rm ML},d=3}_N(\vec{k}),\alpha^{{\rm ML},d=3}_N(\vec{k}')\right]=2i(2\pi)^3\delta^{(3)}(\vec{k}+\vec{k}')
\ee
is still in agreement with \eqref{complcomm}.

While we do not include a further study of the massless case here, we mentioned that the exceptional series representations which we are faced with at this point are reducible. The $\chi^{'+}_{l_\sigma}$ is explicitly decomposed in terms of \emph{discrete series representations} in \cite{Dobrev:1977qv}. In \cite{2007} the massless case is studied in terms of the discrete series representations. Here, by considering the normalization properties of the late-time operators, we are in agreement with these references. 

\subsubsection{Normalized operators corresponding to heavy scalars}
\label{subsec:Normalized operators corresponding to heavy fields}

Heavy scalars with $m^2>\frac{d^2}{4}$ belong to the principal series representations of de Sitter group $SO(d+1,1)$, and for them the properly defined inner product is straight forward
 \be
\label{principinnerproduct} 
\left(\alphaH(\vec{k}),\alphaH(\vec{k})\right)=\frac{1}{\Omega}\int \frac{d^dk}{(2\pi)^d}\langle\alphaH(\vec{k})|\alphaH(\vec{k})\rangle.\ee
There are no shadow transformations of these operators to be obtained with intertwining operators as before. The operators in the principal series representation with mass dependent coefficients are 
 \begin{align}
 \alphaH(\vec{k})&= \NH_{\alpha}\left[-i\frac{\Gamma(i\rho)}{\pi}e^{-\pi\rho/2}a_{\vec{k}}+\frac{1+\coth(\rho\pi)}{\Gamma(1-i\rho)}e^{-\pi\rho/2}a_{-\vec{k}}^\dagger\right]\left(\frac{k}{2}\right)^{-i\rho},\\
 \betaH(\vec{k})&=\NH_{\beta}\left[\frac{1+\coth(\rho\pi)}{\Gamma(1+i\rho)}e^{-\pi\rho/2}a_{\vec{k}}+i\frac{\Gamma(-i\rho)}{\pi}e^{-\pi\rho/2}a^\dagger_{-\vec{k}}\right]\left(\frac{k}{2}\right)^{i\rho}
 \end{align}
 where the convention is such that $\nu^2=\frac{d^2}{4}-m^2=-\rho^2$ and $\rho$ denotes the real positive root, in other words $\nu=i\rho$. Here we are more explicit about the coefficients 
than  \cite{Sengor:2019mbz} and include subdominant contributions as well. Demanding the operators to be normalized with respect to the inner product \eqref{principinnerproduct} implies 
 \begin{align}
     \NH_\alpha&=\sqrt{\rho\pi \sinh(\rho\pi)}e^{-\rho\pi/2},\\
     \NH_\beta&=\sqrt{\rho\pi \sinh(\rho\pi)}e^{\rho\pi/2}
 \end{align}
 where we made use of \eqref{complexgamma} to simplify the complex Gamma functions. This leads to the following normalized operators
\begin{subequations}
\label{normalizedprincp}
\begin{align}
 \alphaHN(\vec{k})&=\sqrt{\rho\pi \sinh(\rho\pi)}\left[-i\frac{\Gamma(i\rho)}{\pi}e^{-\rho\pi}a_{\vec{k}}+\frac{1}{\sinh(\rho\pi)\Gamma(1-i\rho)}a^\dagger_{-\vec{k}}\right]\left(\frac{k}{2}\right)^{-i\rho},\\
 \betaHN(\vec{k})&=\sqrt{\rho\pi \sinh(\rho\pi)}\left[\frac{e^{\rho\pi}}{\sinh(\rho\pi)\Gamma(1+i\rho)}a_{\vec{k}}+i\frac{\Gamma(-i\rho)}{\pi}a^\dagger_{-\vec{k}}\right]\left(\frac{k}{2}\right)^{i\rho}.
 \end{align}
 \end{subequations}
The only nontrivial commutation relation for the operators in the principal series representation is
\begin{align}
\label{princpcommu}
\left[\betaHN(\vec{k}),\alphaHN(\vec{k}')\right]&=-2\sinh(\rho\pi)(2\pi)^d\delta^{(d)}(\vec{k}+\vec{k}').
\end{align}

\subsection{Two-point functions}
\label{sec:two point functions from the late-time operators}
For two given operators $\Ocal_1(\vec{k})$, $\Ocal_2(\vec{k})$ the two-point function is defined as
\be \label{twopoint def} \langle\Ocal_1(\vec{k})\Ocal_2(\vec{k}')\rangle\equiv \langle0|\Ocal_1(\vec{k})\Ocal_2(\vec{k}')|0\rangle.
\ee
Since the structure of the late-time operators are in terms of annihilation and creation operators, it is quite straight forward to compute the two-point functions. If we choose our conventions for the action of creation and annihilation operators as follows 
\begin{subequations}
    \label{creation conventions}
    \begin{align}
   a_{\vec{k}}|0\rangle &=0\\
   a^\dagger_{\vec{k}}|0\rangle&=|\vec{k}\rangle,
    \end{align}
\end{subequations}
then the commutator of the annihilation and creation operators fixes
\be 
\langle \vec{k}|\vec{k}'\rangle=(2\pi)^d\delta^{(d)}(\vec{k}-\vec{k}').
\ee
This will set our convention both for this section and section \ref{sec:Two-point functions via canonical quantization}. 

\subsubsection{Two-point functions for light scalars}
\label{subsubsec:From the late-time boundary operators_compl}
Using the normalized complementary series operators \eqref{normalizedcomp} with the definition \eqref{twopoint def} for the two-point function, 
 we obtain the following two-point functions
\begin{subequations}
\begin{align} 
\langle 
\alphaLN(\vec{k})\alphaLN(\vec{k}')\rangle &=2^\nu k^{-2\nu} \; (2\pi)^d\delta^{(d)}(\vec{k}+\vec{k}'),
\\
\langle \betaLN(\vec{k})\betaLN(\vec{k}')\rangle&=\frac{ k^{2\nu}}{2^\nu}\frac{1+i\cot(\pi\nu)}{1-i\cot(\pi\nu)} \; (2\pi)^d\delta^{(d)}(\vec{k}+\vec{k}'),
\\
\langle \alphaLN(\vec{k})\betaLN(\vec{k}')\rangle&=-i \; (2\pi)^d\delta^{(d)}(\vec{k}+\vec{k}'),
\\
\langle \betaLN(\vec{k})\alphaLN(\vec{k}')\rangle&=i \; \frac{1+i\cot(\pi\nu)}{1-i\cot(\pi\nu)} \; (2\pi)^d\delta^{(d)}(\vec{k}+\vec{k}').
\end{align}
	\label{compnonshadowsector2pt}
\end{subequations}
For completeness we list the two-point functions that involve the shadow sector in appendix \ref{appendix:shadowtwopoint}.

Notice that apart from the Dirac delta function, the cross correlators $\langle \alphaLN(\vec{k})\betaLN(\vec{k}')\rangle$ and $\langle \betaLN(\vec{k})\alphaLN(\vec{k}')\rangle$ are constant in momentum space. When converted to position space these give rise to $\delta^{(d)}\left(\vec{x}-\vec{y}\right)$, a Dirac delta in position space, which are noted as \emph{contact terms}. Reference \cite{Anninos:2023lin} points out further properties and the uniqueness of these contact terms for dS holography as opposed to what is expected in AdS holography, with focus on two spacetime dimensions. We will see such contact terms in sections \ref{subsubsec:From the late-time boundary operators_massless} for discrete series operators and \ref{subsubsec:From the late-time boundary operators_Principal} in the case of principal series late-time operators. However more care must be given whether this statement on contact terms can be carried further to the late-time limit of discrete series field two-point functions \cite{Anninos:2023lin}. 

\subsubsection{Two-point functions for massless scalars}
\label{subsubsec:From the late-time boundary operators_massless}

Focusing only on the discrete series late-time operators of section \ref{subsec:Normalized operators corresponding to massless fields}
\begin{subequations}
\begin{align} \label{alphaMN_4d}\alpha^{M,d=3}_N(\vec{k})&=-i2^{3/4}\left[a_{\vec{k}}-a^\dagger_{-\vec{k}}\right]k^{-3/2}\\
\beta^{M,d=3}_N(\vec{k})&=2^{-3/4}\left[a_{\vec{k}}+a^\dagger_{-\vec{k}}\right]k^{3/2}\end{align}
\end{subequations}
in four spacetime dimensions, we obtain the following two-point functions
\begin{subequations}
    \label{discrete2pt}
\begin{align} 
\langle 
\alpha^{M,d=3}_N(\vec{k})\alpha^{M,d=3}_N(\vec{k}')\rangle &=2^{3/2}k^{-3}(2\pi)^3\delta^{(3)}(\vec{k}+\vec{k}'),
\\
\langle \beta^{M,d=3}_N(\vec{k})\beta^{M,d=3}_N(\vec{k}')\rangle&=2^{3/2}k^3(2\pi)^3\delta^{(3)}(\vec{k}+\vec{k}'),
\\
\langle \alpha^{M,d=3}_N(\vec{k})\beta^{M,d=3}_N(\vec{k}')\rangle&=-i(2\pi)^3\delta^{(3)}(\vec{k}+\vec{k}'),
\\
\langle \beta^{M,d=3}_N(\vec{k})\alpha^{M,d=3}_N(\vec{k}')\rangle&=i(2\pi)^3\delta^{(3)}(\vec{k}+\vec{k}').
\end{align}
\end{subequations}

\subsubsection{Two-point functions for heavy scalars}
\label{subsubsec:From the late-time boundary operators_Principal}
From section \ref{subsec:Normalized operators corresponding to heavy fields}, the normalized principal series operators are \eqref{normalizedprincp}
\begin{align}\label{normalizedprincpsec4}
\alphaHN(\vec{k})&=\sqrt{\rho\pi \sinh(\rho\pi)}\left[-i\frac{\Gamma(i\rho)}{\pi}e^{-\rho\pi}a_{\vec{k}}+\frac{1}{\sinh(\rho\pi)\Gamma(1-i\rho)}a^\dagger_{-\vec{k}}\right]\left(\frac{k}{2}\right)^{-i\rho},\\
 \betaHN(\vec{k})&=\sqrt{\rho\pi \sinh(\rho\pi)}\left[\frac{e^{\rho\pi}}{\sinh(\rho\pi)\Gamma(1+i\rho)}a_{\vec{k}}+i\frac{\Gamma(-i\rho)}{\pi}a^\dagger_{-\vec{k}}\right]\left(\frac{k}{2}\right)^{i\rho}.
 \end{align}
For principal series representations, there is no concept of shadow transformation, so the list of two-point functions will be shorter in this case. From definition \eqref{twopoint def}, a straightforward 
calculation gives
\begin{subequations}
	\label{principalsection2point}
	\begin{align}
	\langle\alphaHN(\vec{k})\alphaHN(\vec{k}')\rangle&=-\frac{\Gamma(1+i\rho)}{\Gamma(1-i\rho)}e^{-\rho\pi}   \left(\frac{k}{2}\right)^{-2i\rho} \; (2\pi)^d\delta^{(d)}(\vec{k}+\vec{k}') ,
\\
\nn&=-e^{2i\gamma_\rho-\rho\pi}  \left(\frac{k}{2}\right)^{-2i\rho} \;  (2\pi)^d\delta^{(d)}(\vec{k}+\vec{k}')
\\
\langle\betaHN(\vec{k})\betaHN(\vec{k}')\rangle &= - \frac{\Gamma(1-i\rho)}{\Gamma(1+i\rho)}e^{\rho\pi} \left(\frac{k}{2}\right)^{2i\rho} 
 \; (2\pi)^d\delta^{(d)}(\vec{k}+\vec{k}') 
\\
\nn&=-e^{-2i\gamma_\rho+\rho\pi}  \left(\frac{k}{2}\right)^{2i\rho} \; (2\pi)^d\delta^{(d)}(\vec{k}+\vec{k}')
\\
\langle\alphaHN(\vec{k})\betaHN(\vec{k}')\rangle&=e^{-\rho\pi}(2\pi)^d\delta^{(d)}(\vec{k}+\vec{k}'),
\\
\langle\betaHN(\vec{k})\alphaHN(\vec{k}')\rangle&=e^{\rho\pi}(2\pi)^d\delta^{(d)}(\vec{k}+\vec{k}').
	\end{align}
\end{subequations}

\section{Two-point functions via canonical quantization}
\label{sec:Two-point functions via canonical quantization}
\subsection{Review of canonical quantization}
\label{subsec:Review of canonical quantization}
In canonical quantization (see e.g. \cite{parker_toms_2009} or \cite{Birrell:1982ix}) 
one expands the field in terms of mode functions $F_{\vec{k}}(\vec{x},\eta)$,  annihilation  $a_{\vec{k}}$ and creation  $a^\dagger_{\vec{k}}$ operators as
\be
\label{phidec_conventionfixed}
\phi(\vec{x},\eta) = \int \frac{d^dk}{(2\pi)^{d/2}}\left\{F_{\vec{k}}(\vec{x},\eta)a_{\vec{k}}+F^*_{\vec{k}}(\vec{x},\eta)a^{\dagger}_{\vec{k}}\right\}.
\ee
The annihilation operator  is such that
\be 
a_{\vec{k}}|0\rangle =0, ~\forall \vec{k}.
\label{eq_annihilation}
\ee
and the creation operator such that
\be 
a^\dagger_{\vec{k}}|0\rangle=|\vec{k}\rangle,
\label{eq_creation}
\ee
The annihilation and creation operators satisfy the commutation relation
	\be
\label{anncomm} 
\comm{a_{\vec{k}}}{a^\dagger_{\vec{k}'}} =(2\pi)^d\delta^{(d)}(\vec{k}-\vec{k}').
\ee
while the states $|\vec{k}\rangle$ satisfy
\be
 \langle \vec{k}|\vec{k}'\rangle=(2\pi)^d\delta^{(d)}(\vec{k}-\vec{k}').
\label{k_states_normalization}
\ee

 The mode functions are normalized with respect to the Klein-Gordon inner product
	\begin{align}
 \label{KGinner} 
\nn
\Klein{F_{\vec{k}}}{F_{\vec{k}'}}
&\equiv -i\int d^dx\sqrt{|g|}g^{0\mu} \left(F^*_{\vec{k}}\partial_\mu F_{\vec{k}'} - F_{\vec{k}'} \partial_\mu F^*_{\vec{k}} \right),
\\
&=\frac{i}{\left(H|\eta|\right)^{d-1}}\int d^dx\left(F^*_{\vec{k}}\partial_\eta F_{\vec{k}'} - F_{\vec{k}'}\partial_\eta F^*_{\vec{k}}  \right)=\delta^{(d)}(\vec{k}-\vec{k}'),
\end{align}
	where $*$ denotes complex conjugation and the over all minus sign in the first line is associated with our $(-,+,+,...)$ sign convention for the metric.
The mode functions define a basis on which a general solution $h(\vec{x},\eta)$ to the field equations can be expanded as~\cite{parker_toms_2009}
\be
 \label{hexpandedoverbassis} 
h(\vec{x},\eta)=\int d^dk\left[F_{\vec{k}}(\vec{x},\eta)\left(F_{\vec{k}},h\right)-F^*_{\vec{k}}(\vec{x},\eta)\left(F^*_{\vec{k}},h\right)\right].
\ee
Since $h(\vec{x},\eta)=\int d^dy\delta^{(d)}(\vec{x}-\vec{y})h(\vec{y},\eta)$, equation \eqref{hexpandedoverbassis} implies a completeness relation for the mode functions
\be 
\label{completenessforF} 
\frac{1}{(H|\eta|)^{d-1}}\int d^dk\left[F_{\vec{k}}(\vec{x},\eta) \; \partial_\eta F^*_{\vec{k}}(\vec{y},\eta) 
  - F^*_{\vec{k}}(\vec{x},\eta) \; \partial_\eta F_{\vec{k}}(\vec{y},\eta)  \right]
= i\delta^{(d)}(\vec{x}-\vec{y})
\ee
in position space.

Given a specific Lagrangian, there exist a momentum operator conjugate to the field operator
\be
 \label{conjugatemom_general} 
\pi\equiv\frac{\partial\mathcal{L}}{\partial\left(\partial_\eta\phi\right)}
\ee
such that the canonically quantized field and its conjugate momentum satisfy the commutation relation
\be 
\label{canonicalquantizationfieldcomm} 
\comm{\phi(\vec{x},\eta)}{\pi(\vec{y},\eta)} = i\delta^{(d)}(\vec{x}-\vec{y}).
\ee

Starting from the action
	\begin{align} 
\label{canquantact}
\nn 
S&=\int^{\eta_0}_{-\infty(1+i\epsilon)}d\eta\int d^dx \sqrt{-g}\left[-\frac{1}{2}g^{\mu\nu}\partial_\mu\phi\partial_\nu\phi-\frac{1}{2}m^2\phi^2\right]
\\
&=
\frac{1}{2}
  \int d\eta \frac{d^dx}{ \left|H\eta\right|^{d-1}  }\left[\left(\partial_\eta\phi\right)^2
  - \left|\vec{\nabla}\phi\right|^2 - \frac{m^2\phi^2}{\left|H \eta\right|^{2}}\right],
	\end{align}
and using \eqref{conjugatemom_general} gives the conjugate momentum as
	\be 
\pi(\vec{x},\eta)
= 
\frac{\partial_{\eta}\phi}{\left|H\eta\right|^{d-1}}
=
\frac{1}{\left|H\eta\right|^{d-1}}
\int\frac{d^dk}{(2\pi)^{d/2}}\left(\partial_\eta F_{\vec{k}} \; a_{\vec{k}} + \partial_\eta F^*_{\vec{k}} \; a^\dagger_{\vec{k}}\right).
\label{conjmomentum}
\ee
	
In what follows below we will work in Fourier space and define the Fourier modes for the canonical field and its conjugate momentum via
\begin{align}
	\label{fourierfield}\phi(\vec{x},\eta)&=\int \frac{d^dk}{(2\pi)^{d}} e^{i\vec{k}\cdot\vec{x}}  \varphi_{\vec{k}}(\eta)
,
\\
	\label{fourierpi} \pi(\vec{x},\eta)&=\int \frac{d^dk}{(2\pi)^{d}}e^{i\vec{k}\cdot\vec{x}}\pi_{\vec{k}}(\eta)
.
\end{align}
	The commutation relation in Fourier space then become
	\be 
\label{fourierspacecomm_canonical} 
\comm{\varphi_{\vec{k}}(\eta)}{\pi_{\vec{k}'}(\eta)} 
=
i(2\pi)^d\delta^{(d)}(\vec{k}+\vec{k}'),
\ee
which is compatible with \eqref{canonicalquantizationfieldcomm}.
	
On de Sitter spacetimes the mode functions take the form 
\be 
F_{\vec{k}}=\Ncal_{\vec{k}} \; \mathcal{F}_{\vec{k}}(\eta) \; e^{i\vec{k}\cdot\vec{x}},
\label{F_Ncal}
\ee
	where, $\Ncal_{\vec{k}}$ is the normalization and $\mathcal{F}_{\vec{k}}(\eta)$ are solutions to the field equations with appropriate initial conditions.
 In our case this is the Bunch Davies initial condition. 
In both the case of light and and the case of heavy scalars, the normalization with respect to the Klein-Gordon inner product gives $\Ncal_{\vec{k}}=\frac{1}{2\sqrt{2}}\left(\frac{H}{2\pi}\right)^{\frac{d-1}{2}}$. 
The difference in the two cases is in the expression for $\mathcal{F}_{\vec{k}}(\eta)$ and the Wronskian of the functions involved. For light scalars 
\be
\label{BDmodes_light}
 \mathcal{F}^L_{\vec{k}}(\eta)=|\eta|^{\frac{d}{2}}H^{(1)}_{\nu}(k|\eta|),~~\nu^2=\frac{d^2}{4}-\frac{m^2}{H^2}
\ee
where $\nu$ is 
real\footnote{ $H^{(1)}_{\nu}(k|\eta|)=J_\nu(k|\eta|)+iY_\nu(k|\eta|)$ is the Hankel function of first kind, $J_\nu(k|\eta|)$ and $Y_\nu(k|\eta|)$ are Bessel functions of first and second kind with Wronskian \cite{NIST:DLMF} $$\mathcal{W}\left\{J_\nu(x),Y_\nu(x)\right\}=\frac{2}{\pi x}.$$ With real index $\nu$ and real argument we have $\left(H^{(1)}_\nu(k|\eta|)\right)^*=H^{(2)}_\nu(k|\eta|)$.}. 
In the case of heavy scalars $\nu$ is purely imaginary, $\nu=i\rho$, and the solution that satisfies the Bunch-Davies initial condition, as also discussed in \cite{Sengor:2019mbz}, 
is~\footnote{Here $\tilde{H}^{(1)}_\rho(k|\eta|)=\tilde{J}_\rho(k|\eta|)+i\tilde{Y}_{\rho}(k|\eta|)$. The Wronskian for the solutions $\tilde{J}_\rho$ and $\tilde{Y}_\rho$ is \cite{NIST:DLMF} $$\mathcal{W}\{\tilde{J}_\rho(x),\tilde{Y}_\rho(x)\}=\tilde{J}_\rho(x)\frac{d}{x}\tilde{Y}_\rho(x)-\left(\frac{d}{dx}\tilde{J}_\rho(x)\right)\tilde{Y}_\rho(x)=\frac{2}{\pi x}.$$}  
\be
\label{BDmodes_heavy} 
\mathcal{F}^H_{\vec{k}}(\eta)=|\eta|^{\frac{d}{2}}\tilde{H}^{(1)}_{\rho}(k|\eta|),~~\rho^2=\frac{m^2}{H^2}-\frac{d^2}{4}.
\ee
	
In general the temporal and spatial dependence of the mode functions factorize as
\be 
\label{splitF} 
F_{\vec{k}}(\vec{x},\eta)=f_{\vec{k}}(\eta)e^{i\vec{k}\cdot\vec{x}},
\ee
with $f_{\vec{k}}(\eta)=f_{-\vec{k}}(\eta)\equiv f_k(\eta)$. So we can write out the momentum modes as
\begin{align}
\varphi_{\vec{k}}(\eta) =& (2\pi)^{d/2}\left(f_{k}  a_{\vec{k}}+f^*_k a^\dagger_{-\vec{k}}\right)
\label{varphik_gen}
\\
\pi_{\vec{k}}(\eta)=&  \frac{(2\pi)^{d/2}}{ |H\eta|^{d-1}  }  \left(\partial_\eta f_{k} \; a_{\vec{k}}  +  \partial_\eta f^*_{k} \; a^\dagger_{-\vec{k}}\right).
	\label{pik_gen} 
\end{align}
	The Klein-Gordon inner product \eqref{KGinner} guarantees that \eqref{varphik_gen} and \eqref{pik_gen} satisfy \eqref{fourierspacecomm_canonical}. One could have also canonically normalized the field at step \eqref{canquantact}, however condition \eqref{KGinner} for normalizing the mode functions implicitly takes this into account.
	
We now follow this procedure to obtain the quantized field and momenta,
 and then calculate the two-point functions in the late-time limit of these operators sandwiched between the vacuum state annihilated by the operator $a_{\vec{k}}$.
We will handle the case of light and heavy scalars separately. The identities for complex conjugation and differentiation of Hankel functions continue to hold under series expansion around the origin at least at leading order. We will write down the canonical field and momenta for all times and take the late time limit before proceeding with the calculation of two-point functions.
	
\subsubsection{Canonically quantized field and momenta for light scalars}
\label{subsubsection:Canonically quantized field and momenta for light fields}	
In the case of light scalars, using the definitions \eqref{F_Ncal} and \eqref{splitF} with $\mathcal{N}_{\vec{k}}$ as found above leads to
\be
 \label{canonicallynormalized mode functions_light} 
f^{\rm L}_{\vec{k}}(\eta)=\frac{1}{2\sqrt{2}}\left(\frac{H}{2\pi}\right)^{\frac{d-1}{2}}|\eta|^{\frac{d}{2}}H^{(1)}_\nu(k|\eta|)
\ee
so that the Fourier space modes are
\begin{align}
\varphiL_{\vec{k}} =
\frac{\sqrt{\pi} }{2}H^{\frac{d-1}{2}}|\eta|^{\frac{d}{2}} 
\left\{ H^{(1)}_\nu(k|\eta|)   a_{\vec{k}} +  \left[  H^{(1)}_\nu(k|\eta|) \right]^* a^\dagger_{-\vec{k}}\right\}.
\label{varphiL_def}
\end{align}
 Taking the late time limit by using the expansion \eqref{Hcompl} then leads to
\begin{align}
\varphiLlt_{\vec{k}}(\eta_0) =&
\frac{\sqrt{\pi} }{2}H^{\frac{d-1}{2}}|\eta_0|^{\frac{d}{2}} 
 \bigg\{ 
\left(\frac{k|\eta_0|}{2}\right)^\nu \left[  \frac{1+i\cot(\pi\nu)}{\Gamma(\nu+1)} a_{\vec{k}} + \frac{1-i\cot(\pi\nu)}{\Gamma(\nu+1)}a^\dagger_{-\vec{k}} \right] 
\nonumber
\\
&
-\frac{i\Gamma(\nu)}{\pi} \left(\frac{k|\eta_0|}{2}\right)^{-\nu}  \left(a_{\vec{k}}  -  a^\dagger_{-\vec{k}} \right) 
\bigg\}.
\end{align}
and employing the normalized late-time boundary operators from \eqref{normalizedcomp} gives
\begin{align}
\varphiLlt_{\vec{k}}(\eta_0) =&
\frac{\sqrt{\pi} }{2}H^{\frac{d-1}{2}}|\eta_0|^{\frac{d}{2}} 
 \bigg\{ 
 \frac{\Gamma(\nu)}{\pi} \left(\frac{\sqrt{2}}{ |\eta_0|}\right)^\nu  \alphaLN(\vec{k}) 
+ \frac{ 1-i\cot(\pi\nu) }{  \Gamma(\nu+1)  } \left(\frac{|\eta_0|}{\sqrt{2}}\right)^\nu   \betaLN(\vec{k})
\bigg\}.
\label{varphilt_L_alphabeta}
\end{align}
Turning now to the canonical momentum operators and using \eqref{pik_gen} leads to
\begin{align}
\piLlt_{\vec{k}}(\eta_0)=&  \frac{\sqrt{\pi}}{ 2 H^{\frac{d-1}{2}}  |\eta_0|^{\frac{d}{2}}  }     \bigg\{
\left(\frac{d}{2} - \nu\right)   \frac{i\Gamma(\nu)}{\pi}\left(\frac{k|\eta_0|}{2}\right)^{-\nu} \; \left(a_{\vec{k}} - a^\dagger_{-\vec{k}}  \right)
\nonumber
\\
&
\label{piLlt} -\left(\frac{d}{2} + \nu\right) \left(\frac{k|\eta_0|}{2}\right)^\nu \left[ \frac{1+i\cot(\pi\nu)}{\Gamma(\nu+1)} \; a_{\vec{k}} +\frac{1-i\cot(\pi\nu)}{\Gamma(\nu+1)} \; a^\dagger_{-\vec{k}} \right]
\bigg\},
\end{align}
and in terms of the normalized late-time boundary operators from \eqref{normalizedcomp} gives
\begin{align}
\piLlt_{\vec{k}}(\eta_0)=&  -\frac{\sqrt{\pi}}{ 2 H^{\frac{d-1}{2}}  |\eta_0|^{\frac{d}{2}}  }     \bigg\{
\left(\frac{d}{2} - \nu\right)   \frac{ \Gamma(\nu)}{\pi}\left(\frac{\sqrt{2}}{|\eta_0|}\right)^{\nu} \; \alphaLN(\vec{k})
\nonumber
\\
&
+ \left(\frac{d}{2} + \nu\right) \frac{ 1-i\cot(\pi\nu) }{ \Gamma(\nu+1) }
\left(\frac{|\eta_0|}{\sqrt{2}}\right)^\nu
\betaLN(\vec{k})
\bigg\}.
	\label{ltpicanL}
\end{align}

\subsubsection{Canonically quantized field and momenta for heavy scalars}
	\label{subsubsection:Canonically quantized field and momenta for heavy fields}

In the case of heavy scalars, using the definitions \eqref{F_Ncal} and \eqref{splitF} with $N_{\vec{k}}$ as found above leads to
\be
 \label{canonicallynormalized mode functions_heavy} 
f^{{\rm H}}_{\vec{k}}(\eta)=\frac{1}{2\sqrt{2}}\left(\frac{H}{2\pi}\right)^{\frac{d-1}{2}}|\eta|^{\frac{d}{2}}  \tilde{H}^{(1)}_\rho(k|\eta|)
\ee
so that the Fourier space modes are
\begin{align}
\varphiH_{\vec{k}} =
\frac{\sqrt{\pi} }{2}H^{\frac{d-1}{2}}|\eta|^{\frac{d}{2}} 
\left\{ \tilde{H}^{(1)}_\rho(k|\eta|)   a_{\vec{k}} +  \left[  \tilde{H}^{(1)}_\rho(k|\eta|) \right]^* a^\dagger_{-\vec{k}}\right\}.
\end{align}
 Taking the late time limit by using the expansion \eqref{Hprinc} then leads to
\begin{align}
\varphiHlt_{\vec{k}}(\eta_0) =&
\frac{\sqrt{\pi} }{2}H^{\frac{d-1}{2}}|\eta_0|^{\frac{d}{2}} 
\bigg\{  
 \left[ \frac{e^{\pi\rho/2}}{ \sinh(\pi \rho) \Gamma(1+i\rho)}a_{\vec{k}}
+ i\frac{\Gamma(-i\rho)}{\pi}e^{-\pi\rho/2}   a^\dagger_{-\vec{k}}
\right] \left(\frac{k|\eta_0|}{2}\right)^{i\rho}
\nonumber
\\
&
+  \left[ \frac{ e^{\pi\rho/2} }{\sinh(\pi \rho)\Gamma(1-i\rho) } a^\dagger_{-\vec{k}}
-i\frac{\Gamma(i\rho)}{\pi}e^{-\pi\rho/2} a_{\vec{k}}
\right] \left(\frac{k|\eta_0|}{2}\right)^{-i\rho}
\bigg\}.
\end{align}
and employing the normalized late-time boundary operators from \eqref{normalizedprincp} gives
\begin{align}
\varphiHlt_{\vec{k}}(\eta_0) =& \frac{ H^{\frac{d-1}{2}}}{2  \sqrt{\rho \sinh(\rho\pi)}   }|\eta_0|^{\frac{d}{2}} \left[ e^{\pi\rho/2}|\eta_0|^{-i\rho} \; \alphaHN(\vec{k}) 
+ e^{-\pi\rho/2} |\eta_0|^{i\rho}  \betaHN(\vec{k}) \right]
\label{varphilt_H_alphabeta}
\end{align}
Turning now to the canonical momentum operators and using \eqref{pik_gen} leads to
\begin{align}
\piHlt_{\vec{k}}(\eta_0)=& 
  \frac{\sqrt{\pi H}}{2  |H\eta_0|^{\frac{d}{2}}  }
 \bigg\{
\left(\frac{d}{2}-i\rho\right) \left[i\frac{ e^{-\frac{\rho\pi}{2}}  \Gamma(i\rho)}{\pi}a_{\vec{k}}
-  \frac{ e^{\frac{\rho\pi}{2}}  }{\sinh(\rho\pi)\Gamma(1-i\rho)}a^\dagger_{-\vec{k}}\right]\left(\frac{k|\eta_0|}{2}\right)^{-i\rho}
\nonumber
\\
& 
- \left(\frac{d}{2}+i\rho\right)  \left[\frac{e^{\frac{\rho\pi}{2}}}{\sinh(\rho\pi)\Gamma(1+i\rho)}a_{\vec{k}}
+i\frac{ e^{\frac{\rho\pi}{2}}  \Gamma(-i\rho)}{\pi}a^\dagger_{-\vec{k}}\right]\left(\frac{k|\eta_0|}{2}\right)^{i\rho}
\bigg\}.
\end{align}
and in terms of the normalized late-time boundary operators from \eqref{normalizedprincp} gives
\begin{align}
\piHlt_{\vec{k}}(\eta_0)=& 
 - \frac{\sqrt{H} |H\eta_0|^{-\frac{d}{2}} }{2 \sqrt{\rho \sinh(\rho\pi)}  }
 \bigg[
\left(\frac{d}{2}-i\rho\right)|\eta_0|^{-i\rho}e^{\frac{\rho\pi}{2}}\alphaHN(\vec{k})
\nonumber
\\
& \ \ \ \
+ \left(\frac{d}{2}+i\rho\right)|\eta_0|^{i\rho} e^{-\frac{\rho\pi}{2}}  \betaHN(\vec{k})
\bigg].
	\label{pikLT_H_alphabeta}
\end{align}
	
\subsection{Two-point functions}

Up to this point we have gathered all the ingredients we need to compare the methods of evaluating correlators via the wavefunction, the late-time operators and canonical quantization. By doing so, we hope to gain more insight into what the two different late-time operators $\alpha$ and $\beta$ physically represent. We will once again work out the case of light and heavy scalars individually. Our definition for the two-point function is that mentioned in equation \eqref{twopoint def}, sandwiching the operators of interest in between the vacuum states.

\label{subsec:Two-point functions in canonical quantization}

\subsubsection{Two-point functions for light scalars}
\label{subsec:canonicalquantlatetimetwopoint_light}
By concentrating on the organization of canonically quantized field and conjugate momentum modes as given in \eqref{varphilt_L_alphabeta} and \eqref{ltpicanL}, we can immediately observe the relation for the late-time two-point functions between the formulations of canonical quantization and late-time operators.
		\begin{align}
\nn
\langle\varphiLlt_{\vec{k}}\varphiLlt_{\vec{k}'}\rangle
=& \frac{\pi}{4}H^{d-1}|\eta_0|^d\Bigg\{
\frac{2^\nu\Gamma^2(\nu)}{\pi^2 |\eta_0|^{2\nu}}\langle\alphaLN(\vec{k})\alphaLN(\vec{k}')\rangle
+\frac{(1-i\cot(\nu\pi))^2}{2^\nu\Gamma^2(1+\nu)}|\eta_0|^{2\nu}\langle\betaLN(\vec{k})\betaLN(\vec{k}')\rangle
\\
&+\frac{1-i\cot(\nu\pi)}{\nu\pi}\left[\langle\betaLN(\vec{k})\alphaLN(\vec{k}')\rangle +\langle\alphaLN(\vec{k})\betaLN(\vec{k}')\rangle \right]
\Bigg\}
		\label{ltphipiK}
\\
=&
\frac{ \pi (2\pi)^d }{4}H^{d-1}|\eta_0|^d\Bigg\{
\frac{\Gamma^2(\nu)}{\pi^2} \left(\frac{ k|\eta_0|}{2} \right)^{-2\nu}
+\frac{(1+\cot^2(\nu\pi))}{\Gamma^2(1+\nu)} \left(\frac{ k|\eta_0|}{2} \right)^{2\nu}
\nonumber
\\
&
-\frac{2}{\nu\pi} \cot(\pi\nu) 
\Bigg\} \delta^{(d)}(\vec{k}+\vec{k}')
	\label{twoptcanquant_light_k}
		\end{align}
where in the last relation we have used  the properties of the two-point functions of the late-time  boundary operators~\eqref{compnonshadowsector2pt}. 
Similarly we find
\begin{align}
\nn
\langle\piLlt_{\vec{k}}\piLlt_{\vec{k}'}\rangle =& \frac{\pi}{4}\frac{1}{|\eta_0|^dH^{d-1}}\Bigg\{
\frac{1-i\cot(\nu\pi)}{\nu\pi}\left(\frac{d}{2}+\nu\right)\left(\frac{d}{2}-\nu\right)
 \bigg[\langle\betaLN(\vec{k})\alphaLN(\vec{k}')\rangle
\\
&
 +\langle\alphaLN(\vec{k})\betaLN(\vec{k}')\rangle \bigg]
+\frac{2^\nu\Gamma^2(\nu)}{\pi^2 |\eta_0|^{2\nu}}\left(\frac{d}{2}-\nu\right)^2 \langle\alphaLN(\vec{k})\alphaLN(\vec{k}')\rangle
\nn
\\
\label{lightpipi_latetimops}&
+\frac{(1- i\cot(\nu\pi))^2|\eta_0|^{2\nu}}{2^\nu\Gamma^2(\nu+1)}\left(\frac{d}{2}+\nu\right)^2 \langle\betaLN(\vec{k})\betaLN(\vec{k}')\rangle
\Bigg\}.
	    \end{align}
and
\begin{align}
\nn
\langle\piLlt_{\vec{k}}\piLlt_{\vec{k}'}\rangle
 =& \frac{\pi}{4}\frac{(2\pi)^d}{|\eta_0|^dH^{d-1}}\Bigg[
\frac{\Gamma^2(\nu)}{\pi^2}\left(\frac{d}{2}-\nu\right)^2\left(\frac{k|\eta_0|}{2}\right)^{-2\nu}
+ \frac{1+\cot^2(\nu\pi)}{\Gamma^2(\nu+1)}\left(\frac{d}{2}+\nu\right)^2\left(\frac{k|\eta_0|}{2}\right)^{2\nu}
\\
&
-\frac{2\cot(\nu\pi)}{\nu\pi}\left(\frac{d}{2}+\nu\right)\left(\frac{d}{2}-\nu\right)
\Bigg]
\delta^{(d)}(\vec{k}+\vec{k}').\label{canLpipiK}
	    \end{align}
for the equivalent relations in the case of the canonical momentum two-point functions.

The above relations establish the equality of 
$\langle\varphiLlt_{\vec{k}}\varphiLlt_{\vec{k}'}\rangle =\langle\Phi_{\vec{k}}\Phi_{\vec{k}'}\rangle$
and
$\langle\piLlt_{\vec{k}}\piLlt_{\vec{k}'}\rangle =\langle\Pi_{\vec{k}}\Pi_{\vec{k}'}\rangle$
and gives the relation between the wavefunction and late-time boundary two-point functions.

\subsubsection{The massless case}
\label{subsubsec:Comparing the operators and latetime profiles compl}
Before moving to the heavy scalars, let us consider the massless case where $\nu=\frac{d}{2}$. 
Among the late-time operators as we saw in section \ref{subsec:Normalized operators corresponding to massless fields}, the $\alpha$ in this case belong to the category of exceptional series representations in any dimension. The normalization for this category works differently than that of complementary series representations. Depending on the number of dimensions under consideration, $\beta$ operators belong to either complementary or exceptional series categories.

As noted before, in four spacetime dimensions, $(d=3)$, which is the case relevant for inflation, exceptional series representations are identified with discrete series \cite{sun2021note}. Thus $\alpha^{M,d+3}$ belongs to the discrete series representations while $\beta^{M,d=3}$ seems to belong to the complementary series representations as identified in \ref{subsec:Normalized operators corresponding to massless fields}.

In canonical quantization picture, looking at the momentum dependence of the late-time two-point functions we see from equations \eqref{twoptcanquant_light_k} and \eqref{canLpipiK} that for the massless case

		\begin{align}
\nn
\langle\varphi^{{\rm ML},lt}_{\vec{k}}\varphi^{{\rm ML},lt}_{\vec{k}'}\rangle
=&
\frac{ \pi (2\pi)^d }{4}H^{d-1}|\eta_0|^d\Bigg\{
\frac{\Gamma^2(\frac{d}{2})}{\pi^2} \left(\frac{ k|\eta_0|}{2} \right)^{-d}
+\frac{(1+\cot^2(\frac{d}{2}\pi))}{\Gamma^2(1+\frac{d}{2})} \left(\frac{ k|\eta_0|}{2} \right)^{d}
\nonumber
\\
&
-\frac{2}{\frac{d}{2}\pi} \cot\left(\frac{d}{2}\pi\right) 
\Bigg\} \delta^{(d)}(\vec{k}+\vec{k}'),
\end{align}
and
\begin{align}
\nn
\langle\pi^{{\rm ML},lt}_{\vec{k}}\pi^{{\rm ML},lt}_{\vec{k}'}\rangle
=& \frac{\pi}{4}\frac{d^2(2\pi)^d}{H^{d-1}}\Bigg[ \frac{1+\cot^2(\frac{d}{2}\pi)}{\Gamma^2(\frac{d}{2}+1)}\left(\frac{k}{2}\right)^{d}\Bigg]
\delta^{(d)}(\vec{k}+\vec{k}').
\end{align}
The momentum dependence of the late-time conjugate momentum in canonical quantization carries only the momentum dependence coming from $\beta^M$. 

In $d=3$, using the results of section \ref{subsubsection:Canonically quantized field and momenta for light fields} we can go further and make the following identification in terms of normalized late-time operators given in equations \eqref{alphaMN} and \eqref{normalizedBetaMasless}
\begin{align}
    \varphi^{{\rm ML},d=3,lt}_{\vec{k}}(\eta_0)&=H|\eta_0|^{\frac{3}{2}}\left\{\frac{2^{1/4}}{3}|\eta_0|^{3/2}\beta^{{\rm ML},d=3}_N(\vec{k})+|\eta_0|^{-3/2}\frac{1}{2^{5/4}}\alpha^{{\rm ML},d=3}_N(\vec{k})\right\},
\\
    \pi^{{\rm ML},d=3,lt }_{\vec{k}}(\eta_0)&=-\frac{1}{H}2^{1/4}\beta^{{\rm ML},d=3}_N(\vec{k}).
\end{align}
In fact from \eqref{piLlt} or \eqref{ltpicanL}, in the massless case with $\nu=\frac{d}{2}$, the contribution from $\alpha$ operators to $\piLlt$ vanish in any dimension. 

Thus we obtain
\begin{subequations}
\label{masslesstwopoints}
\begin{align}
\nn
    \langle 
\varphi^{{\rm ML},d=3,lt}_{\vec{k}}\varphi^{{\rm ML},d=3,lt}_{\vec{k}'}\rangle&=H^2|\eta_0|^3\Bigg\{\frac{|\eta_0|^{-3}}{2^{5/2}}\langle\alpha^{{\rm ML},d=3,lt}_N\alpha^{{\rm ML},d=3,lt}_N\rangle+\frac{\sqrt{2}}{9}|\eta_0|^3
 \langle\beta^{{\rm ML},d=3,lt}_N\beta^{{\rm ML},d=3,lt}_N\rangle
\\
    &~~~~+\frac{1}{6}\left[\langle\beta^{{\rm ML},d=3}_N\alpha^{{\rm ML},d=3}_N\rangle+\langle\alpha^{{\rm ML},d=3}_N\beta^{{\rm ML},d=3}_N\rangle\right]\Bigg\}
\\
    \langle \pi^{{\rm ML},lt d=3}_{\vec{k}}\pi^{{\rm ML},lt d=3}_{\vec{k}'}\rangle&=\frac{\sqrt{2}}{H^2}\langle\beta^{{\rm ML},d=3}_N\beta^{{\rm ML},d=3}_N\rangle.
\end{align}
\end{subequations}

Note that in the case of a massless field in $d=3$, only the $\beta_N$ autocorrelator contributes to the canonical quantized conjugate momentum late-time two point function (which we know matches the wavefunction two-point function). Moreover this late-time two-point function does not have any $\eta_0$ dependence.

	\subsubsection{Two-point functions for heavy scalars}
	\label{subsec:canonicalquantlatetimetwopoint_heavy}
Concentrating on the organization \eqref{varphilt_H_alphabeta} and \eqref{pikLT_H_alphabeta} in terms of the late-time operators and making use of the appropriate commutation relation in this case, we obtain
\begin{align}
\nn
\langle\varphiHlt_{\vec{k}}\varphiHlt_{\vec{k}'}\rangle
=&
\frac{H^{d-1}|\eta_0|^d}{4\rho \sinh(\rho\pi)}\Bigg[
 |\eta_0|^{-2i\rho}e^{\rho\pi}\langle\alphaHN(\vec{k})\alphaHN(\vec{k'})\rangle
+ |\eta_0|^{2i\rho}e^{-\rho\pi} \langle\betaHN(\vec{k})\betaHN(\vec{k'})\rangle  
\\
&
+\langle\alphaHN(\vec{k})\betaHN(\vec{k'})\rangle+\langle\betaHN(\vec{k})\alphaHN(\vec{k'})\rangle\Bigg]
\label{canphiphi_alphabeta}
\end{align}
and using \eqref{principalsection2point}
 \begin{align}
\nn
\langle\varphiHlt_{\vec{k}}\varphiHlt_{\vec{k}'}\rangle
=&
\frac{ (2\pi)^d H^{d-1}|\eta_0|^{d} }{4\rho \sinh(\rho\pi)}\Bigg[
2\cosh(\rho\pi)
\\
&
-e^{-2i\gamma_\rho}\left(\frac{k|\eta_0|}{2}\right)^{2i\rho}-e^{2i\gamma_\rho}\left(\frac{k|\eta_0|}{2}\right)^{-2i\rho}\Bigg]
\delta^{(d)}(\vec{k}+\vec{k}')
        \label{canphiphi_H}
        \end{align}
Similarly for the canonical momentum we obtain
\begin{align}
\nn\langle\piHlt_{\vec{k}}\piHlt_{\vec{k}'}\rangle=&\frac{|\eta_0|^{-d}}{4\rho \sinh(\rho\pi) H^{d-1}}\Bigg\{\left(\frac{d}{2}+i\rho\right)^2|\eta_0|^{2i\rho}e^{-\rho\pi}\langle\betaHN(\vec{k})\betaHN(\vec{k'})\rangle\\
\nn&+\left(\frac{d}{2}-i\rho\right)^2|\eta_0|^{-2i\rho}e^{\rho\pi}\langle\alphaHN(\vec{k})\alphaHN(\vec{k'})\rangle
\\
&
+\left(\frac{d^2}{4}+\rho^2\right)\left[ \langle\alphaHN(\vec{k})\betaHN(\vec{k'})\rangle+\langle\betaHN(\vec{k})\alphaHN(\vec{k'})\rangle\right] \Bigg\}.
\label{canpipi_alphabeta}
\end{align}
and using \eqref{principalsection2point}
	\begin{align}
    \nn
\langle\piHlt_{\vec{k}}\piHlt_{\vec{k}'}\rangle
=&
\frac{(2\pi)^d}{4\rho \sinh(\rho\pi)|\eta_0|^d H^{d-1}}
\Bigg[2 \left(\frac{d^2}{4}+\rho^2\right)\cosh(\rho\pi)
-\left(\frac{d}{2}+i\rho\right)^2e^{-2i\gamma_\rho}\left(\frac{k|\eta_0|}{2}\right)^{2i\rho}
\\
& 
-\left(\frac{d}{2}-i\rho\right)^2e^{2i\gamma_\rho}\left(\frac{k|\eta_0|}{2}\right)^{-2i\rho}\Bigg]
\delta^{(d)}(\vec{k}+\vec{k}')
.
	\label{canpipi_H}
	\end{align}
The above relations establish the equality of 
$\langle\varphiHlt_{\vec{k}}\varphiHlt_{\vec{k}'}\rangle =\langle\PhiH_{\vec{k}}\PhiH_{\vec{k}'}\rangle$
through \eqref{2pointheavyPhi} and \eqref{princp2ptPi} and
$\langle\piHlt_{\vec{k}}\piHlt_{\vec{k}'}\rangle =\langle\PiH_{\vec{k}}\PiH_{\vec{k}'}\rangle$
and gives the relation between the wavefunction and late-time boundary two-point functions.

\section{Conclusions}
\label{sec:conclusions}
\subsection{Summary of results}
In an earlier work, we identified operators at the late-time boundary of de Sitter that correspond to free massive scalar fields and recognized them in terms of the unitary irreducible representations of $SO(d+1,1)$, the symmetry group of $d+1$ dimensional de Sitter \cite{Sengor:2019mbz}. The unitary irreducible representations of $SO(d+1,1)$ are labelled by their spin and scaling weight. By checking the well-defined inner product for the operators we obtained, we confirmed that light scalars correspond to complementary series representations and have real scaling weights, which lead to real scaling dimensions. Likewise, heavy scalars correspond to principal series representations with purely imaginary scaling weights and complex scaling dimensions. In this current work, we also gathered evidence towards identifying a late-time operator corresponding to discrete series representations from the late-time behaviour of a massless scalar on $dS_4$. 
The Bunch Davies initial conditions (imposing the field to be well behaved at early times), 
allowed us to talk about two different operators at the late-time boundary for a given massive scalar field. We denoted the operator with the 
lower scaling dimension by $\alpha$ and the operator with the higher scaling dimension by $\beta$. 

The fact that two operators exist at the boundary, and that the principal series representations with complex scaling dimensions are allowed and are among the unitary representations, are all 
features unique to scalar fields on de Sitter. These features are different than what applies to scalar fields on Anti de Sitter where the boundary conditions allow for only one operator for given mass 
and a complex scaling dimension is not allowed (see \cite{Dias_2010}, \cite{maldacena2014gaugegravity} for a review). In considering quantum fields on de Sitter and Anti de Sitter, differences such as these arise because different symmetry groups are at work. The symmetry group of Anti de Sitter in $d+1$ dimensions is the group $SO(d,2)$. Contrary to $SO(d+1,1)$, the group $SO(d,2)$ includes time translation. Due to the presence of time translation symmetry, one can talk about positive energy eigenstates. This puts a lower bound on the mass of a scalar field on Anti de Sitter \cite{Breitenlohner:1982bm,Breitenlohner:1982jf}. Such a bound does not apply to scalar fields on de Sitter because the group $SO(d+1,1)$ does not involve time translation symmetry, there is no global timelike Killing vector for de Sitter. 
Moreover the presence of time translation invariance requires scaling dimensions for fields on Anti de Sitter to be real, and this is the concept of unitarity that applies there.
 On the other hand a more natural concept of unitarity for de Sitter is associated with the unitarity of representations and this works into the definition of the inner product. The well-defined inner product 
is different for each of the categories of discrete, complementary and principal series representations 
and this is because the reality of the scaling weight for discrete and complementary series representations needs to be treated carefully. 
While the inner product for the principal series is straightforward, the inner product for the other categories involves a shadow transformation, as explicitly demonstarated for complementary series in \cite{Sengor:2019mbz} and for discrete series here. 
The correlators and the wavefunction can be simplified to mimic the form of a CFT partition function  
in the case of light scalars, but it involves subtleties in the case of heavy scalars \cite{Isono2020}. 
These are unintuitive features when compared to scalar fields on AdS. Yet these features exist and should warn us about reaching too quick conclusions regarding
field theory on de Sitter.\footnote{In fact section 5D of \cite{Dobrev:1977qv} begins with a warning that only some of the representations of $SO(d+1,1)$ can be continued to a special class of the unitary representations of universal covering of $SO(d,2)$, and cites \cite{Lepowsky1973AlgebraicRO}.}

The aim of this work was to create a link between the late-time field profiles $\{\Phi_{\vec{k}},\Pi_{\vec{k}}\}$, the late-time canonical quantization fields $\{\varphilt_{\vec{k}},\pi^{lt}_{\vec{k}}\}$ and the normalized late-time operators $\{\alphaN(\vec{k}),\betaN(\vec{k})\}$ in $k$-space, via the computation of two-point functions. We calculated the late-time two-point functions of $\{\Phi_{\vec{k}},\Pi_{\vec{k}}\}$ 
using the wavefunction method. We computed the late-time wavefunction for complementary series as well as massless scalars and principal series representations in sections \ref{subsec:The Bunch Davies wavefunction for light scalars}, \ref{subsec:The Bunch Davies wavefunction for heavy scalars} respectively, on Poincaré coordinates for de Sitter and without resorting to analytic continuation from the EAdS wavefunction. 
The two-point functions were then calculated in sections \ref{subsubsec:The complementary series wavefunction} and \ref{subsubsec:From the principal series wavefunction} for light and heavy scalars respectively. The calculation of the two-point functions from the late-time operators were quite straightforward, and was carried out in section \ref{subsubsec:From the late-time boundary operators_compl} for complementary series operators, in section \ref{subsubsec:From the late-time boundary operators_massless} for discrete series operators and in section \ref{subsubsec:From the late-time boundary operators_Principal} for principal series operators. The canonical quantization picture in section \ref{sec:Two-point functions via canonical quantization} makes it apparent that the field and conjugate momentum operators at late-times are linear combinations of the late-time operators $\alphaN$ and $\betaN$ in both cases of light and heavy scalars. Comparing the three sets of calculations for the two-point functions, schematically both light field two-point functions ( with $\mu=\nu$ as given by equations \eqref{ltphipiK}, \eqref{lightpipi_latetimops}) and heavy field two-point functions (with $\mu=i\rho$, as given by equations \eqref{canphiphi_alphabeta}, \eqref{canpipi_alphabeta}) have a similar form 
\begin{align}
  \langle\Phi_{\vec{k}}\Phi_{\vec{k}'}\rangle=\langle\varphi^{lt}_{\vec{k}}\varphi^{lt}_{\vec{k}'}\rangle=&H^{d-1}|\eta_0|^d\Bigg\{c_1|\eta_0|^{-2\mu}\langle\alphaN\alphaN\rangle+c_2|\eta_0|^{2\mu}\langle\betaN\betaN\rangle+c_3\left[\langle\alphaN\betaN\rangle+\langle\betaN\alphaN\rangle\right]\Bigg\}\\
  \nn\langle\Pi_{\vec{k}}\Pi_{\vec{k}'}\rangle=\langle\pi^{lt}_{\vec{k}}\pi^{lt}_{\vec{k}'}\rangle=&\frac{1}{|\eta_0|^dH^{d-1}}\Bigg\{c_1|\eta_0|^{-2\mu}\left(\frac{d}{2}-\mu\right)^2\langle\alphaN\alphaN\rangle+c_2|\eta_0|^{2\mu}\left(\frac{d}{2}+\mu\right)^2\langle\betaN\betaN\rangle\\
  &+c_3\left(\frac{d}{2}-\mu\right)\left(\frac{d}{2}+\mu\right)\left[\langle \alphaN\betaN\rangle+\langle\betaN\alphaN\rangle\right]\Bigg\} 
  \end{align}
with differences only in the coefficients $c_i$. In the case of a massless scalar only $\beta_N$ contributes to the conjugate momentum and some of these coefficients are zero as discussed in \ref{subsubsec:Comparing the operators and latetime profiles compl}.

A recent proposal is to introduce mixed boundary conditions to simplify the principal series field two-point function \cite{Isono2020}. This proposal only focuses on field two-point functions. In the complementary series two-point function, one can neglect the presence of $\betaLN$ which we identified in sections \ref{subsubsec:The complementary series wavefunction} and \ref{subsec:canonicalquantlatetimetwopoint_light} arguing that the coefficient $|\eta_0|^{d+2\nu}$ with $\nu$ being real and positive is negligible at late-times. This is the simplification that \cite{Isono2020} make, leading to a two-point function that can easily be recognized as a two-point function that would arise in a CFT. In the case of principal series two-point function however, the contribution from what we identify as the late-time operator $\betaHN$ cannot be neglected in the late-time limit, giving the principal series two-point function a puzzling appearance and leading to the mixed boundary conditions proposal as a possible remedy. As stated in \cite{Isono2020}, the wavefunction with the mixed boundary conditions is related to the wavefunction with the Bunch-Davies boundary conditions, the latter of which is the one we use here, by a Legendre transformation. Here we would like to point out that in principle both of the late-time operators $\alphaN$ and $\betaN$ contribute in both cases of complementary series and principal series two-point functions.

In the main body of our work we implement the late-time limit from the start. However in appendix \ref{app:latetime at the end} we redo all our calculations in the bulk and leave the implementation of the late-time limit at the end. We get the same results in both methods and confirm that the order of the limits do not cause problems.

\subsection{Outlook}
\label{subsec:outlook}	
	 In this work we have given a list of scalar two-point functions at the late-time boundary of de Sitter to accompany the list of scalar late-time boundary operators whose form and scaling dimensions were identified previously in \cite{Sengor:2019mbz} and extended our list which consisted of principal and complementary series operators to include the discrete series operators. Our hope is that those familiar with Euclidean Conformal Field Theories can suggest possible dual currents to these late-time boundary operators we have identified. Of course the list should be enlarged to include gauge fields and fermions for completeness. We leave this task for future work. 

Our aim of constructing the late-time operators seems to bring along a method of calculating late-time correlation functions. Inclusion of gauge fields which belong to discrete series representations and fields with more general spin will allow us to further develop this method and explore its merits. On a parallel line, it is also tempting to consider interactions and explore the Clebsch-Gordon coefficients between different operators. There has been recent advances along this line, especially with focus on the principal series representations, in perturbative quantum field theory \cite{DiPietro:2021sjt} and non-perturbative conformal bootstrap programme \cite{Hogervorst:2021uvp}, that touch upon the implications of bulk unitarity of interactions for the boundary. We leave all these points for future work.

With this and the preceding work \cite{Sengor:2019mbz} we consider the concept of unitarity at the late-time boundary and focus on how representation theory manifests itself at the late-time boundary. On a complementary route, building bulk fields from the unitary irreducible representations \cite{2006princ},\cite{2007compl} have also been investigated. Moreover there is a lot of ongoing investigation on the concept of unitarity for bulk interactions, such as \cite{2021timeev}, \cite{2021optical} to cite very recent work along this line. It remains to be understood further how the bulk and boundary approaches to unitarity complement each other.   

Heavy scalars on de Sitter seem to be less studied than their light counterparts. However, we have seen that they have a lot of interesting properties. 
Thus we would like to conclude by pointing out growing interest in principal series representations. The scalar members of the principal series representations, that is scalars on de Sitter whose mass is heavy with respect to the Hubble scale, play a role in cosmological applications. In \cite{Armendariz_Picon_2018} they have been considered in the context of perturbative decay of the inflaton field so as to question if the end products of such decays take over before inflation even ends. In \cite{Arkani-Hamed:2015bza} they are recognized to have interesting imprints on the primordial correlation functions, this is the oscillatory nature that the imaginary exponent brings along, which we also obtained in \eqref{2pointheavyPhi} for the late-time field profile, in \eqref{principalsection2point} for the late-time operators and in \eqref{canphiphi_H} in the canonical quantization picture.
 In the DFF model~\cite{Anous:2020nxu},  a quantum mechanical model, the principal series representations in two dimensions and how these representations can be accommodated within that model
 have been discussed. An interesting feature of two dimensions is that here the Hamiltonian is among the symmetry generators, contrary to the case in higher dimensions. Moreover the principal series representations can also appear on two dimensional AdS once charged fields are considered \cite{Anninos:2019oka}. The principal series representations also appear in the context of flatspace holography \cite{Donnay:2020guq,Pasterski:2017kqt}. On the other hand, there do exist CFTs with operators that have complex scaling dimensions\footnote{We thank the participants of the King's College London Theoretical Physics online Journal Club for drawing our attention to some of these examples.}. One example is the proposal of \emph{Complex CFTs} \cite{Gorbenko_2018}. However, the case of interest here is with operators such that the real part of the scaling dimension is close to marginality, $\frac{d+1}{2}$ in our convention of counting dimensions, where as the real part of the dimension of our principal series operators are $\frac{d}{2}$. Another example is the fishnet model \cite{G_rdo_an_2016} where both purely imaginary and complex scaling dimensions can arise in certain values of the coupling parameters \cite{Cavagli__2020}. It is an open question if any of the known CFT's with complex scaling dimensions could capture the heavy scalars on de Sitter.

\acknowledgments

We would like to sincerely thank Dionysios Anninos, Tarek Anous, Taha Ayfer, Armelle Bajat, Paolo Benincasa, Nicolas Boulanger, Alejandro Cabo-Bizet, Andrea Cavagli$\grave{a}$, Claudia de Rham, Cem Eröncel, Atsushi Higuchi, Vasileios Letsios, Ben Pethybridge, Bayram Tekin, Ayngaran Thavanesan and Andrew Tolley for insightful discussions. GŞ acknowledges support by the European Union's Horizon 2020 research and innovation programme under the Marie Sk\l{}odowska-Curie grant agreement No 840709-SymAcc in the first half and from the European Structural and Investment  Funds  and  the  Czech  Ministry  of  Education,  Youth  and  Sports  (MSMT)  
(Project  CoGraDS  -CZ.02.1.01/0.0/0.0/15003/0000437) in the second half of this work.
CS acknowledges support from the European Structural and Investment  Funds  and  the  Czech  Ministry  of  Education,  Youth  and  Sports  (MSMT)  
(Project  CoGraDS  -CZ.02.1.01/0.0/0.0/15003/0000437).

\appendix

\section{Two-point functions in the shadow sector}
\label{appendix:shadowtwopoint}
For completeness, we list here the two-point functions that involve the shadow sector in the case of complementary series representations. We start by giving a brief review of the shadow transformation by considering how it acts on $\alphaLN$. 

Our operators $\alphaLN$ and $\betaLN$ belong to the function space $\mathcal{C}_{\chi}$ with $\chi=\{0,c\}$ denoting the label of representations. The label $c$ is the scaling weight, which is $c=-\nu$ for the operator $\alphaLN$, and $c=\nu$ for the operator $\betaLN$ as established in \cite{Sengor:2019mbz}. The shadow operators belong to the function space $\mathcal{C}_{\tilde{\chi}}$ where for the case of scalars $\tilde{\chi}=\{0,-c\}$. The shadow transformation is employed via the intertwining operator, whose explicit form for scalars is
\be \label{intert} G_{\chi}: \mathcal{C}_{\tilde{\chi}}\to \mathcal{C}_{\chi}~\text{for}~ c<0,~G^+_{\{0,c\}}(k)=\left(\frac{k^2}{2}\right)^c\ee
where the superscript $+$ denotes the choice for normalization. This operator also has the inverse
\be \label{inverseinter} G_{\tilde{\chi}}: \mathcal{C}_{\chi}\to\mathcal{C}_{\tilde{\chi}}~\text{for}~c>0,~G^+_{\{0,\tilde{c}\}}(k)=\left(\frac{k^2}{2}\right)^{\tilde{c}}.
\ee
For the late-time operator
\be
\alphaL(\vec{k})=-\frac{i}{\pi}\Gamma(\nu)N_\alpha
 \left[a_{\vec{k}}-a^\dagger_{-\vec{k}}\right]\left(\frac{k}{2}\right)^{-\nu},~~c=-\nu,\ee
 its shadow $\tilde{\alpha}^L$ is extracted from
 \begin{align}
  \nn   \alpha^L(\vec{k})=G_{\{0,-\nu\}}(k)\tilde{\alpha}^L(\vec{k})\\
  -\frac{i}{\pi}\Gamma(\nu)N_\alpha
 \left[a_{\vec{k}}-a^\dagger_{-\vec{k}}\right]\left(\frac{k}{2}\right)^{-\nu}=\left(\frac{k^2}{2}\right)^{-\nu}\tilde{\alpha}^L(\vec{k}).
 \end{align}

 In the end, the normalized shadow late-time operators are
\begin{align}
\alphaLNt(\vec{k})&=-i2^{-\nu/2}k^\nu\left[a_{\vec{k}}-a^\dagger_{-\vec{k}}\right],\\
\betaLNt(\vec{k})&=2^{\nu/2}k^{-\nu}\left[\frac{1+i\cot(\pi\nu)}{1-i\cot(\pi\nu)}a_{\vec{k}}+a^\dagger_{-\vec{k}}\right].
\end{align}
The two-point functions of the shadow sector are given by
\begin{align}
\langle\alphaLNt(\vec{k})\alphaLNt(\vec{k}')\rangle&=2^{-\nu}k^{2\nu}(2\pi)^d\delta^{(d)}(\vec{k}+\vec{k}'),\\
\langle\betaLNt(\vec{k})\betaLNt(\vec{k}')\rangle&=2^\nu k^{-2\nu}\frac{1+i\cot(\pi\nu)}{1-i\cot(\pi\nu)}(2\pi)^d\delta(\vec{k}+\vec{k}')\\
\langle \alphaLNt(\vec{k})\betaLNt(\vec{k}')\rangle&=-i(2\pi)^d\delta^{(d)}(\vec{k}+\vec{k}')\\
\langle\betaLNt(\vec{k})\alphaLNt(\vec{k}')\rangle&=i\frac{1+i\cot(\pi\nu)}{1-i\cot(\pi\nu)}(2\pi)^d\delta^{(d)}(\vec{k}+\vec{k}'),
\end{align}
while the two-point functions involving a shadow and a non-shadow operator are given by
\begin{align}
\langle \alphaLN(\vec{k})\alphaLNt(\vec{k}')\rangle&=(2\pi)^d\delta^{(d)}(\vec{k}+\vec{k}')\\
\langle \alphaLN(\vec{k})\betaLNt(\vec{k}')\rangle&=-i2^{\nu} k^{-2\nu}(2\pi)^d\delta^{(d)}(\vec{k}+\vec{k}')\\
\langle \alphaLNt(\vec{k})\betaLN(\vec{k}')\rangle&=-i2^{-\nu}k^{2\nu}(2\pi)^d\delta^{(d)}(\vec{k}+\vec{k}').
\end{align}

\section{Taking the late-time limit after computing the two-point functions} 
\label{app:latetime at the end}
In the main text we take the late-time limit from the start of the computations, making the 
 identification of the late-time operators manifest from the start. This later feature is most apparent when we discuss canonical quantization in section \ref{sec:Two-point functions via canonical quantization}. In this appendix we carry on the two-point calculations in wavefunction and canonical quantization formalisms in the bulk and take the late-time limit at the end, to make sure that our approach of taking the late-time limit from the beginning does not cause us to loose any information. The results match with the approach of the main text. And important step along the way that allows one to still have a track able appearance of the late-time operators is to rewrite the annihilation and creation operators in terms of the late-time operators. We will organize this appendix similar to the way we have organized the main text.

\subsection{Revisiting the wavefunction method}
\label{subapp:wavefunction}
Whether we take the late-time limit or not, the wavefunction has the general form as a functional of time and late-time profile
\be
\label{Psiform_bulk} 
\Psi_{BD}\left[\Phi,\eta\right]=\Ncal(\eta)\exp\left[-\frac{1}{2}\int\frac{d^dk}{(2\pi)^d}\Pcal(k,|\eta|)\Phi_{\vec{k}}\Phi_{-\vec{k}}\right].
\ee
similar to the one given in \eqref{Psiform}. The main difference is that in \eqref{Psiform_bulk} the time $\eta$ is the bulk time while in \eqref{Psiform} it is the late-time value $\eta_0\to 0$. The calculations of section \ref{sec:two-point functions} still hold, leading to 
\begin{subequations}
\label{wavefunction bulk 2pt form}
    \begin{align} 
\label{field2point_form}
\langle \Phi_{\vec{k}}\Phi_{\vec{k}'}\rangle&=\frac{(2\pi)^d}{\Pcal^*(k,|\eta|)+\Pcal(k,|\eta|)}\delta^{(d)}(\vec{k}+\vec{k}'),\\
\langle\Pi_{\vec{k}}\Pi_{\vec{k}'}\rangle&=(2\pi)^d\frac{\Pcal^*(k,|\eta|)\Pcal(k,|\eta|)}{\Pcal^*(k,|\eta|)+\Pcal(k,|\eta|)}\delta^{(d)}(\vec{k}+\vec{k}').
\end{align}
\end{subequations}
To calculate these two-point functions for any given time $\eta$, we just need to compare \eqref{Psiform_bulk} with the semiclassical approximation
\begin{align}
    \Psi_{BD}\left[\Phi,\eta\right]\sim e^{iS_{onshell}[\Phi,\eta]},
\end{align}
to read off $\Pcal[k,|\eta|]$. We will do this category by category.

\subsubsection{Complementary series wavefunction and bulk two-point functions}
\label{subapp:compl buk wavefunction 2pt}
We already obtained the onshell action in the bulk for complementary series light fields in section \ref{subsec:The Bunch Davies wavefunction for light scalars} as
\be \label{eq:onshell bulk compl} S^{compl}_{onshell}=-\frac{1}{2H^{d-1}}\int \frac{d^dk}{(2\pi)^{d}}|\eta_0|^{-d}|\Phi_{\vec{k}}|^2\left\{-|\eta_0|\frac{d}{d|\eta|}\left[\frac{|\eta|^{d/2}H^{(1)}_\nu(k|\eta|)}{|\eta_0|^{d/2}H^{(1)}_\nu(k|\eta_0|)}\right]_{|\eta|=|\eta_0|}\right\}.\ee
For emphasis, let us call the $\Pcal$ in equation \eqref{Psiform_bulk} as $\Pcal_{bulk}$. and for convenience let us define $z\equiv k|\eta|$. Then we read off $\Pcal$ for the complementary series case as
\begin{subequations}
    \begin{align}
        \Pcal^{compl}_{bulk}(z)&=-\frac{ik^dz_0^{-d}}{H^{d-1}}\left\{\frac{d}{2}+\frac{z_0}{H^{(1)}_\nu(z_0)}\left[\frac{d}{dz}H^{(1)}_\nu(z)\right]_{z=z_0}\right\}\\
\label{eqnapp:Pbcompl}     &= -\frac{i}{|\eta_0|^dH^{d-1}}\left\{\frac{d}{2}+\frac{z_0H^{(2)}_\nu(z_0)}{|J_\nu(z_0)|^2+|Y_\nu(z_0)|^2}\left[\frac{d}{dz}H^{(1)}_\nu(z)\right]_{z=z_0}\right\}  
    \end{align}
\end{subequations}
where we have made use of the fact that for complementary series $\left(H^{(1)}_\nu(z)\right)^*=H^{(2)}_\nu(z)$, since $\nu$ is real. Then the terms that go into the two-point function calculations are easy to obtain
\begin{subequations}
    \begin{align}
 \Pcal^{compl}_{bulk}(k,|\eta_0|)+\left(\Pcal^{compl}_{bulk}(k,|\eta_0|)\right)^*&=\frac{4}{\pi |\eta_0|^dH^{d-1}}\frac{1}{|J_\nu(k|\eta_0|)|^2+|Y_\nu(k|\eta_0|)|^2}\\
 \Pcal^{compl}_{bulk}(k,|\eta_0|)\left(\Pcal^{compl}_{bulk}(k,|\eta_0|)\right)^*&=\frac{z_0^2}{4}\frac{\left|2\left[\frac{dJ_\nu}{dz}\right]_{z_0}+\frac{d}{z_0}J_\nu(z_0)\right|^2+\left|2\left[\frac{dY_\nu}{dz}\right]_{z_0}+Y_\nu(z_0)\right|^2}{|\eta_0|^{2d}H^{2(d-1)}\left[|J_\nu(z_0)|^2+|Y_\nu(z_0)|^2\right]}
    \end{align}
\end{subequations}
Even though $\eta_0$ makes an appearance in these expressions, since we have not yet taken the limit $\eta_0\to 0$ is can be any finite value within the bulk. In the bulk we arrive at the following results for the two-point functions
\begin{subequations}
    \begin{align}
\langle\PhiL_{\vec{k}}\PhiL_{\vec{k}'}\rangle&=\frac{\pi|\eta_0|^dH^{d-1}}{4}\left[|J_\nu(k|\eta_0|)|^2+|Y_\nu(k|\eta_0|)|^2\right](2\pi)^d\delta^{(d)}\left(\vec{k}+\vec{k}'\right)\
\\
\langle\PiL_{\vec{k}}\PiL_{\vec{k}'}\rangle&=k^2|\eta_0|^2\pi\frac{(2\pi)^d\delta^{(d)}\left(\vec{k}+\vec{k}'\right)}{16|\eta_0|^dH^{d-1}}\left[\left|2\left[\frac{dJ_\nu}{dz}\right]_{z_0}+\frac{d}{z_0}J_\nu(z_0)\right|^2+\left|2\left[\frac{dY_\nu}{dz}\right]_{z_0}+Y_\nu(z_0)\right|^2\right].
    \end{align}
\label{gen_compl_wavefunction}
\end{subequations}

\subsubsection{Principal series wavefunction and bulk two-point functions}
\label{subapp:princ buk wavefunction 2pt}
Now let us carry on the same procedure to obtain bulk principal series two-point functions. In section \ref{subsec:The Bunch Davies wavefunction for heavy scalars} we obtained the following onshell action in the bulk for principal series fields 
\begin{align}
    S^{{\rm Prin}}_{onshell}=-\frac{1}{2H^{d-1}}\int \frac{d^dk}{(2\pi)^{d}}|\eta_0|^{-d}|\Phi_{\vec{k}}|^2\left\{-|\eta_0|\frac{d}{d|\eta|}\left[\frac{|\eta|^{d/2}\tilde{H}^{(1)}_\rho(k|\eta|)}{|\eta_0|^{d/2}\tilde{H}^{(1)}_\rho(k|\eta_0|)}\right]_{|\eta|=|\eta_0|}\right\}.
\end{align}
We will again define $z\equiv k|\eta|$ for convenience. This time $\Pcal$ in equation \eqref{Psiform_bulk} reads
\begin{align}
    \Pcal^{princ}_{bulk}(z)&=-\frac{i}{H^{d-1}}\frac{z_0^{-d+1}}{k^{-d}}\frac{d}{dz}\left[\frac{z^{d/2}\tilde{H}^{(1)}_\rho(z)}{z_0^{d/2}\tilde{H}^{(1)}_\rho(z_0)}\right]_{z=z_0}\\
    &=-\frac{i}{H^{d-1}|\eta_0|^d}
    \left\{\frac{d}{2}+\frac{z_0}{\tilde{H}^{(1)}_\rho(z_0)}\left[\frac{d}{dz}\tilde{H}^{(1)}_\rho(z)\right]_{z=z_0}
    \right\}.
\end{align}
Given the definitions in the main text, we make use of the fact that $\left(\tilde{H}^{(1)}_\rho(z)\right)^*=\tilde{H}^{(2)}_\rho(z)$ and arrive at 
\begin{subequations}
    \begin{align}
        \Pcal^{princ}_{bulk}(k,|\eta_0|)+\left(\Pcal^{princ}_{bulk}(k,|\eta_0|)\right)^* &= \frac{4}{\pi H^{d-1}|\eta_0|^d}\frac{1}{\tilde{H}^{(1)}_\rho(k|\eta_0|)\tilde{H}^{(2)}_\rho(k|\eta_0|)},
\\ 
  \Pcal^{princ}_{bulk}(k,|\eta_0|)\left(\Pcal^{princ}_{bulk}(k,|\eta_0|)\right)^* &= \frac{z_0^2}{4}\frac{\left|2\left[\frac{d\tilde{J}_\rho}{dz}\right]_{z_0}+\frac{d}{z_0}\tilde{J}_\rho(z_0)\right|^2+\left|2\left[\frac{d\tilde{Y}_\rho}{dz}\right]_{z_0}+\tilde{Y}_\rho(z_0)\right|^2}{|\eta_0|^{2d}H^{2(d-1)}\left[|\tilde{J}_\rho(z_0)|^2+|\tilde{Y}_\rho(z_0)|^2\right]}. 
    \end{align}
\end{subequations}
Via equations \eqref{wavefunction bulk 2pt form} we arrive at 
\begin{subequations}
\begin{align}
\langle\PhiH_{\vec{k}}\PhiH_{\vec{k}'}\rangle
 &=
\frac{\pi|\eta_0|^d H^{d-1}}{4}
\left[
 \left|\tilde{J}_\rho\right|^2 
+ \left|\tilde{Y}_\rho\right|^2 
\right]_{z=z_0} \; (2\pi)^d\delta^{(d)}\left(\vec{k}+\vec{k}'\right),
\label{gen_princip_wavefunction_Phi}
\\
\langle\PiH_{\vec{k}}\PiH_{\vec{k}'}\rangle&=
\frac{  \pi (2\pi)^d }{ 4 H^{d-1} |\eta_0|^{d}  }  
\left[
\left| \frac{d}{2}  \tilde{J}_\rho  +   z_0 \frac{d\tilde{J}_\rho}{dz} \right|^2  
+  \left| \frac{d}{2}  \tilde{Y}_\rho  + z_0 \frac{d\tilde{Y}_\rho}{dz}\right|^2
\right]_{z=z_0}
\delta^{(d)}(\vec{k}+\vec{k}')
\label{gen_princip_wavefunction_Pi}
\end{align}
\label{gen_princip_wavefunction}
\end{subequations}
where to remind the reader,
 $\tilde{J}_\rho(z) = \sech\left(\frac{\rho\pi}{2}\right) \Real\left[J_{i\rho}(z)\right]$ 
and $\tilde{Y}_\rho(z)= \sech\left(\frac{\rho\pi}{2}\right)\Real\left[Y_{i\rho}(z)\right]$~\cite{doi:10.1137/0521055}.

\subsection{Revisiting canonical quantization}
In this section we derive the two-point functions using canonical quantization but without a series expansion from the outset. The starting point is relations
\eqref{varphik_gen} and \eqref{pik_gen}, which express the $k$-space field and conjugate momentum operators in terms of creation and annihilation operators, $a_{\vec{k}}^\dagger$ and $a_{\vec{k}}$. We then express the 
creation/annihlation operators in terms of the  late-time operators, and form the two-point correlation functions using the  two-point functions  of the late-time operators.  While the step of passing to
the late-time operators is unnecessary, as one can find the two-point correlation functions directly 
using \eqref{eq_annihilation}, \eqref{eq_creation} and \eqref{k_states_normalization}, we choose to do that in order to make the connection with the late-time operators explicit.

We perfom the necessary steps separately for the complementary and for the principal series below.

\subsubsection{Complementary series}
We start from the relations \eqref{normalizedcomp} which express the normalized late-time operators $\alphaLN(\vec{k})$ and $\betaLN(\vec{k})$ in terms of $a_{\vec{k}}^\dagger$ and $a_{\vec{k}}$. 
Inverting those relations leads to
\begin{align}
a_{\vec{k}} =& \frac{i}{2} \left(\frac{k}{\sqrt{2}} \right)^\nu  \left(1 -  i \cot\pi\nu\right) \alphaLN 
+  \frac{1}{2}\left(\frac{k}{\sqrt{2}} \right)^{-\nu}    \left(1 - i  \cot\pi\nu\right)  \betaLN,
\\
a^{\dagger}_{-\vec{k}} =&  -\frac{i}{2} \left(\frac{k}{\sqrt{2}} \right)^\nu    \left(1 + i  \cot\pi\nu\right) \alphaLN 
 +  \frac{1}{2}\left(\frac{k}{\sqrt{2}} \right)^{-\nu}    \left(1 - i   \cot\pi\nu\right)  \betaLN.
\end{align}
We plug the above two equations  into \eqref{varphik_gen} and \eqref{pik_gen}, and after some algebra we find
\begin{align}
\varphiL_{\vec{k}} =& \frac{\sqrt{\pi}}{2}
H^{\frac{d-1}{2}}
|\eta|^{\frac{d}{2}} 
\bigg\{  
  \left(\frac{k}{\sqrt{2}} \right)^\nu  \left[ \cot(\pi\nu) J_\nu  - Y_\nu  \right]\alphaLN
+  \left(\frac{k}{\sqrt{2}} \right)^{-\nu}   \left(1 - i  \cot\pi\nu\right)  J_\nu \betaLN
\bigg\}.
\end{align}
and
\begin{align}
\piL_{\vec{k}}=&  -\frac{z\sqrt{\pi}}{4H^{\frac{d-1}{2}} |\eta|^{d/2}  }  
\bigg\{
  \left(\frac{k}{\sqrt{2}} \right)^\nu    \bigg[ \cot(\pi\nu)   \left(  2 \frac{dJ_\nu}{dz} + \frac{d}{z}  J_{\nu} \right)
- \left( 2 \frac{dY_\nu}{dz}  + \frac{d}{z}  Y_{\nu} \right) \bigg] \alphaLN
\nonumber
\\
&
+ \left(\frac{k}{\sqrt{2}} \right)^{-\nu}   \left(1 - i  \cot\pi\nu\right)  \left[ 2 \frac{dJ_\nu}{dz}  + \frac{d}{z}  J_\nu \right] \betaLN
\bigg\}
\end{align}
Taking the two-point functions and using the complementary series late-time operator two-point functions  \eqref{compnonshadowsector2pt} followed by further algebraic manipulations  leads to the exact relations
\begin{align}
\langle \varphiL_{\vec{k}} \varphiL_{\vec{k}'}\rangle =&
\langle \PhiL_{\vec{k}} \PhiL_{\vec{k}'}\rangle 
\\
\langle \piL_{\vec{k}} \piL_{\vec{k}'}\rangle =& \langle \PiL_{\vec{k}} \PiL_{\vec{k}'}\rangle 
\end{align}
with the RHS given by \eqref{gen_compl_wavefunction}.

\subsubsection{Principal series}
We start from the relations \eqref{normalizedprincp} which express the normalized late-time operators $\alphaHN(\vec{k})$ and $\betaHN(\vec{k})$ in terms of $a_{\vec{k}}^\dagger$ and $a_{\vec{k}}$. 
Inverting those relations leads to
\begin{align}
a_{\vec{k}} =&  \frac{1}{ 2 \sqrt{\pi \rho \sinh(\pi \rho)} }  \left[
 \Gamma(1- i \rho) 
 \left(\frac{k}{2}\right)^{i \rho} \alphaHN 
+     \Gamma(1+ i \rho)  \left(\frac{k}{2}\right)^{-i \rho} \betaHN,
\right]
\\
a^{\dagger}_{-\vec{k}} =&  
 \frac{1}{ 2\sqrt{\pi \rho  \sinh(\pi \rho) } }  \left[
 e^{ \pi \rho } \Gamma(1- i \rho)   \left(\frac{k}{2}\right)^{i \rho} 
 \alphaHN 
+  e^{-\pi \rho} \Gamma(1+i\rho)   \left(\frac{k}{2}\right)^{-i \rho}  \betaHN.
\right]
\end{align}
We plug the above two equations  into \eqref{varphik_gen} and \eqref{pik_gen}, and after some algebra we find
\begin{align}
\varphiH_{\vec{k}} =& 
 \frac{1}{4 \sqrt{\rho \sinh(\pi \rho)}  }H^{\frac{d-1}{2}}|\eta|^{\frac{d}{2}}  
\bigg\{
  \left( \tilde{H}^{(1)}_\rho  + \tilde{H}^{(2)}_\rho  e^{ \pi \rho } \right) \Gamma(1- i \rho) \left(\frac{k}{2}\right)^{i \rho} \alphaHN
\nonumber
\\
&
+  \left(  \tilde{H}^{(1)}_\rho +   \tilde{H}^{(2)}_\rho   e^{-\pi \rho}  \right)\Gamma(1+ i \rho)   \left(\frac{k}{2}\right)^{-i \rho}  \betaHN
\bigg\}.
\end{align}
and
\begin{align}
\piH_{\vec{k}} =&  -\frac{ z  }{ 8H^{\frac{d-1}{2}} |\eta|^{d/2}    \sqrt{ \rho \sinh(\pi \rho)}   }  
\bigg\{
\nonumber
\\
&
 \left[ \frac{d}{z}   \tilde{H}^{(1)}_\rho(z)  + 2\frac{d\tilde{H}^{(1)}_\rho}{dz}
+   e^{ \pi \rho }   \left( \frac{d}{z}   \tilde{H}^{(2)}_\rho(z)  + 2\frac{d\tilde{H}^{(2)}_\rho}{dz} \right)
\right] 
\Gamma(1- i \rho) \left(\frac{k}{2}\right)^{i \rho} \alphaHN 
\nonumber
\\
&
 +  \left[ \frac{d}{z}   \tilde{H}^{(1)}_\rho(z)  + 2\frac{d\tilde{H}^{(1)}_\rho}{dz}
  + e^{-\pi \rho}   \left( \frac{d}{z}   \tilde{H}^{(2)}_\rho(z)  + 2\frac{d\tilde{H}^{(2)}_\rho}{dz}\right) \right]  \Gamma(1+i\rho)   \left(\frac{k}{2}\right)^{-i \rho}  \betaHN
\bigg\}
\end{align}
Taking the two-point functions and using the principal series late-time operator two-point functions \eqref{principalsection2point} followed by further algebraic manipulations  leads to the exact relations
\begin{align}
\langle \varphiH_{\vec{k}} \varphiH_{\vec{k}'}\rangle =& \langle \PhiH_{\vec{k}} \PhiH_{\vec{k}'}\rangle 
\\
\langle \piH_{\vec{k}} \piH_{\vec{k}'}\rangle =& \langle \PiH_{\vec{k}} \PiH_{\vec{k}'}\rangle 
\end{align}
with the RHS given by \eqref{gen_princip_wavefunction}.

\subsection{Series expansions of the two-point functions} 
\subsubsection{Late-time complementery series two-point functions from the bulk}
Starting from \eqref{gen_compl_wavefunction} we expand the Bessel functions according to whether $\nu$ is integer (with $\nu=0$ a special case) or non-integer real number.
Specifically for any $\nu\in \mathcal{R}$ and $\nu\ge 0$, we have that
\begin{equation}
 J_\nu(z) = \left(\frac{z}{2}\right)^{\nu}\sum_{k=0}^{\infty}(-1)^{k}\frac{ \left( \frac{z^2}{4}\right)^{k}}{k! \, \Gamma\left(\nu+k+1\right) }  
\label{Bessel_J}
\end{equation}
while the series expansion of the Bessel function of the 2nd kind is 
\begin{align}
Y_\nu(z)  =&
 \cot(\pi \nu) \left(\frac{z}{2}\right)^{\nu}\sum_{k=0}^{\infty}(-1)^{k}\frac{ \left( \frac{z^2}{4}\right)^{k}}{k! \, \Gamma\left(\nu+k+1\right) }  
 -  \csc(\pi\nu) 
\left(\frac{z}{2}\right)^{-\nu}\sum_{k=0}^{\infty}(-1)^{k}\frac{ \left( \frac{z^2}{4}\right)^{k}}{k! \, \Gamma\left(-\nu+k+1\right) } 
\label{Series_Y}
\\
=& - \frac{  \Gamma(- \nu)   \cos(\pi \nu)}{\pi}
  \left(\frac{z}{2}\right)^{\nu}
 \bigg[
1
- \frac{   \frac{z^2}{4}}{ \nu+1 }  
+ \ldots
\bigg]
 -  \frac{ \Gamma(\nu) }{ \pi }
\left(\frac{z}{2}\right)^{-\nu}
\bigg[
1
- \frac{ \left( \frac{z^2}{4}\right)}{1-\nu} 
+ \ldots
\bigg]
\label{Series_Y_Mathematica}
\end{align}
when $\nu>0$ is not an integer, and 
\begin{align}
 Y_n(z) 
 =& 
\frac{2}{\pi}  \ln\left(\frac{z}{2}\right) J_n(z)
- \frac{1}{\pi} \left(\frac{z}{2}\right)^{n} \sum_{k=0}^{\infty} (-1)^{k}\frac{ \left[ \psi\left(n+k+1\right) +  \psi\left(k+1\right) \right]
}{k!(n+k)!} \left( \frac{z^2}{4}\right)^{k} 
 \nonumber
\\
&
- \frac{1}{\pi} \left(\frac{z}{2}\right)^{-n} \sum_{k=0}^{n-1}  \frac{(n-k-1)!}{k!} \left( \frac{z^2}{4}\right)^{k}
\label{Series_Y_n}
\end{align}
when  $\nu = n$ is a non-zero positive integer.  Of special case is  $n=0$ so that
\begin{align}
Y_0(z)
 =&
\frac{2}{\pi}  \left( \gamma_E +  \ln \frac{z}{2} \right) J_0(z)
- \frac{2}{\pi} \sum_{k=0}^{\infty} (-1)^{k}\frac{H_k}{(k!)^2} \left( \frac{z^2}{4}\right)^{k} 
\label{Series_Y_0}
\end{align}
where $\psi(k+1) = -\gamma_E + H_k$, with $\gamma_E$ being the Euler-Mascheroni constant and $H_k = \sum_{m=1}^k \frac{1}{m}$  the $k$-th Harmonic number,
and $H_0=0$. 

With these expansions, the general formulas \eqref{gen_compl_wavefunction} result in \eqref{2pointlightPhi} and \eqref{compl2ptPi} for the non-integer case, 
while when $\nu = n$ is a non-zero integer we find
\begin{align}
\langle \PhiL_{\vec{k}}\PhiL_{\vec{k}'}\rangle=&  = \langle \varphiL_{\vec{k}}\varphiL_{\vec{k}'}\rangle  
= \frac{ \pi |\eta_0|^d H^{d-1}   (2\pi)^d}{ 4 }  
 \bigg\{
\frac{1}{ n!^2  }
 \left[ 1 + \frac{4}{\pi^2}   \left( \gamma_E -  \frac{1}{2} H_n + \ln \frac{z}{2} \right)^2 \right] \left(\frac{z}{2}\right)^{2n}
\nonumber
\\
&
-  \frac{4}{n\pi^2}  \left[ \gamma_E -  \frac{1}{2} H_n + \ln \frac{z}{2} \right] 
+ \frac{   (n-1)!^2 }{\pi^2} \left(\frac{z}{2}\right)^{-2n}   
 \bigg\}
\delta^{(d)}(\vec{k}+\vec{k}'),
\\
\langle\PiL_{\vec{k}}\PiL_{\vec{k}'}\rangle=& \langle\piL_{\vec{k}}\piL_{\vec{k}'}\rangle = \frac{  \pi (2\pi)^d  }{4|\eta_0|^{d} H^{d-1}}  \bigg\{
 \frac{ (\frac{d}{2} + n )^2}{ n!^2   } \left(\frac{z}{2}\right)^{2n}  
+ \frac{4}{\pi^2}  \frac{1}{ n!^2 } \left[ 1 + \left(\frac{d}{2} + n \right) \left( \gamma_E - \frac{1}{2} H_n +  \ln \frac{z}{2} \right) \right]^2 \left(\frac{z}{2}\right)^{2n}   
\nonumber
\\
&
+ \frac{4}{n\pi^2}   \left(n - \frac{d}{2}  \right)\left[ 1 + \left(\frac{d}{2} + n \right) \left( \gamma_E - \frac{1}{2} H_n +  \ln \frac{z}{2} \right) \right] 
\nonumber
\\
&
+  \frac{(n-1)!^2}{\pi^2} \left(n - \frac{d}{2}  \right)^2 \left(\frac{z}{2}\right)^{-2n}  
\bigg\} \delta^{(d)}(\vec{k}+\vec{k}'),
\end{align}
and if $n= 0$ we get
\begin{align}
\langle \PhiL_{\vec{k}}\PhiL_{\vec{k}'}\rangle=&  \langle \varphiL_{\vec{k}}\varphiL_{\vec{k}'}\rangle   =\frac{ \pi |\eta_0|^d H^{d-1}   (2\pi)^d}{ 4 }  
 \bigg\{
 1 +   \frac{4}{\pi^2} \left( \gamma_E +  \ln \frac{z}{2} \right)^2 
\nonumber
\\
&
+ \left[ -1  +  \frac{8}{\pi^2} \left( \gamma_E +  \ln \frac{z}{2} \right) \left(1 - \gamma_E - \ln \frac{z}{2} \right) \right] \frac{z^2}{4  } 
 \bigg\}
\delta^{(d)}(\vec{k}+\vec{k}'),
\\
\langle\PiL_{\vec{k}}\PiL_{\vec{k}'}\rangle=&  \langle\piL_{\vec{k}}\piL_{\vec{k}'}\rangle  = \frac{  \pi  (2\pi)^d  }{16|\eta_0|^{d} H^{d-1}}  \bigg\{
 d^2 
+ \frac{4}{\pi^2} \left[  2 + d   \left( \gamma_E +  \ln \frac{z}{2} \right) \right]^2
+  \bigg[ d (4-d)
\nonumber
\\
&
+ \frac{8}{\pi^2} \left[  2 + d   \left( \gamma_E +  \ln \frac{z}{2} \right) \right]
 \left[ 2+d - (4+d) \left( \gamma_E +  \ln \frac{z}{2} \right) \right]
\bigg]
 \frac{z^2}{4}  
 \bigg\} \delta^{(d)}(\vec{k}+\vec{k}').
\end{align}

\subsubsection{Late-time principle series two-point functions from the bulk}
In this case we need the series expansion of Bessel functions of imaginary order. With the terms readjusted conveniently we have \cite{NIST:DLMF,Sengor:2019mbz}
\begin{align}
 \tilde{J}_\rho(z) =& 
 \sqrt{\frac{   \tanh\left(\frac{\rho\pi}{2}\right)   }{ 2\pi\rho } }
\bigg[
  e^{-i\gamma_{\rho}}  \left(\frac{z}{2}\right)^{i \rho} \left( 1 - \frac{1}{ 1 + i \rho} \frac{z^2}{4} + \ldots \right)
\nonumber
\\
& 
+ e^{i\gamma_{\rho}}  \left(\frac{z}{2}\right)^{-i \rho}  \left( 1 - \frac{1}{ 1 - i \rho} \frac{z^2}{4} + \ldots \right)
\bigg]
\label{Bessel_J_tilde}
\end{align}
and
\begin{align}
 \tilde{Y}_\rho(z) =& 
 - i \sqrt{\frac{   \coth\left(\frac{\rho\pi}{2}\right)   }{ 2\pi\rho } }
 \bigg\{
e^{-i\gamma_{\rho}}
  \left(\frac{z}{2}\right)^{i\rho}
 \bigg[
1
- \frac{1 }{1 + i \rho} \frac{z^2}{4}
+ \ldots
\bigg]
\nonumber
\\
&
- e^{i\gamma_{\rho}}
\left(\frac{z}{2}\right)^{- i\rho}
\bigg[
1
- \frac{1}{1-i \rho } \frac{z^2}{4}
+ \ldots
\bigg]
\bigg\}
\label{Bessel_Y_tilde}
\end{align}
Plugging these expansions into \eqref{gen_princip_wavefunction_Phi}  and keeping only the leading order we recover \eqref{2pointheavyPhi} and \eqref{canphiphi_H}. Similarly,
plugging these expansions into \eqref{gen_princip_wavefunction_Pi}  and keeping only the leading order we recover  \eqref{princp2ptPi} and \eqref{canpipi_H}.

\bibliography{twopoint_arxiv_v2}
\bibliographystyle{JHEP}	
	\end{document}